\documentclass[showpacs,aps,floatfix,prb,reprint,superscriptaddress]{revtex4-1}
\usepackage{xfrac}
\usepackage{amssymb}
\usepackage{braket}
\usepackage{graphicx}
\usepackage{verbatim}
\usepackage{etoolbox}
\usepackage{amsfonts} 
\usepackage{amsmath} 
\usepackage[utf8]{inputenc} 
\usepackage{mathtools}
\usepackage{soul,xcolor,colortbl}
\usepackage{bm} 
\usepackage{array}
\newcolumntype{C}[1]{>{\centering\arraybackslash}p{#1}}\usepackage{soul}
\definecolor{Gray}{gray}{0.85}

\usepackage{color}  
\begin{document}

\title{Improved predictions of the physical properties of Zn- and Cd-based wide band-gap semiconductors: a validation of the ACBN0  functional \\
}

\author{Priya Gopal}
\author{Marco Fornari}

\email{marco.fornari@cmich.edu}
\affiliation{Department of Physics, Central Michigan University, Mt. Pleasant, MI 48859, USA}
\affiliation{Center for Materials Genomics, Duke University, Durham, NC 27708, USA}

\author{Stefano Curtarolo}
\affiliation{Materials Science, Electrical Engineering, Physics and Chemistry, Duke University, Durham, NC 27708, USA}
\affiliation{Center for Materials Genomics, Duke University, Durham, NC 27708, USA}

\author{Luis A. Agapito}
\author{Laalitha S.I. Liyanage}
\author{Marco Buongiorno Nardelli}
\email{mbn@unt.edu}
\affiliation{Department of Physics, University of North Texas, Denton, TX 76203, USA}
\affiliation{Center for Materials Genomics, Duke University, Durham, NC 27708, USA}

\date{\today}

\begin{abstract}
We study the physical properties of Zn\textit{X} (\textit{X}=O, S, Se, Te) and Cd\textit{X} (\textit{X}=O, S, Se, Te)
in the zinc-blende, rock-salt, and wurtzite structures using the recently developed 
fully \textit{ab initio} pseudo-hybrid Hubbard density functional ACBN0.
We find that both the electronic and vibrational properties of these wide-band gap semiconductors are systematically improved over the PBE values and reproduce closely the  experimental measurements. Similar accuracy is found for the structural parameters, especially the bulk modulus.
ACBN0  results compare well with hybrid functional calculations at a fraction of the computational cost.
\end{abstract}

\maketitle

\section{Introduction}
\label{sec:intro}

Among II-VI semiconductors, oxides and chalcogenides with Cd and Zn are 
the subject of extensive scrutiny for 
their potential applications in spintronics, optoelectronics, and photovoltaics.\cite{ZnX-book}
These applications rely on hetero-junctions, quantum dots, and other nanostructures where 
 the delicate interplay between structural and electronic 
properties must be carefully captured to understand and improve the
device's performance.
Common computational approaches based on density functional theory (DFT), however, fall
short when trying to predict accurate structural and electronic properties, greatly limiting the
development process.

In DFT, most of the calculations involving II-VI semiconductors and oxides use the exchange-correlation (xc) potential
within the local density approximation (LDA) or the generalized gradient approximation (GGA).\cite{LDA,PBE}
While the DFT-LDA/GGA method provides reasonable predictions for structural properties, it fails dramatically in describing the electronic properties, especially the band-gaps, which are underestimated by  50\% or more with respect to experimental values. In ZnO, for instance, the calculated band gap is 0.8 eV,  while the experimental value is 3.44 eV; \cite{Mang_ZnO_SSC1995,Powell_photoemission_ZnO_PRB1972} CdO is described as a semi-metal,\cite{Schelfe_CdO_PRB2006} whereas experimentally rocksalt CdO has an indirect band gap of 0.9 eV. \cite{Demchenko_CdOexpt_PRB2010} Similar discrepancies in the calculated and measured band-gaps are found in the chalcogenides as well.\cite{Vogel_self_interaction_pseudopotentials_II_IV_PRB1996,Ley_valence_band_density_III_V_II_VI_PRB1974}
In the case of Cd- and Zn- oxides and chalcogenides, in addition, the energy of the occupied  \textit{d}-manifold is 
found  about 3 eV higher than the experimental value and lead to  fictitious \textit{d-p} mixing in the valence band.\cite{Vogel_self_interaction_pseudopotentials_II_IV_PRB1996,Ley_valence_band_density_III_V_II_VI_PRB1974}

The limitations mentioned above are a consequence of the lack of self-interaction corrections\cite{PerdewZunger}  and of the derivative discontinuity in the exchange-correlation energy\cite{Perdew83, Sham83, Perdew1982DFT_for_fractional} in all the traditional functionals such as LDA or PBE-GGA.  Clearly, only the inclusion of non-local correlations and non-local Hartree-Fock (HF) exchange can, in part, alleviate these problems. 
A number of approaches  have been proposed to overcome these deficiencies, among which: 
LDA plus self-interaction (LDA-SIC),\cite{PerdewZunger} self-interaction-relaxation correction (SIRC-LDA),\cite{Vogel_self_interaction_pseudopotentials_II_IV_PRB1996} DFT+U,\cite{LDAU,LDAU1}
different versions of the GW approximation,\cite{GW,Fuchs_quasiparticles_KS_PRB2007,Shishkin_scGW_PRB2007} and  hybrid functionals.\cite{HSE, HSE06}
Among these approaches, the ones that have gained more traction in recent years are the DFT+ U\cite{LDAU,LDAU1} and hybrid functionals.\cite{HSE, HSE06}

In DFT+U one introduces  on-site Coulomb and exchange interaction terms, $U$ and $J$,  to account for the localization of \textit{d} states. 
The on-site orbital-dependent parameters $U$ and $J$ correspond to the Coulomb and exchange couplings between electrons of a particular angular momentum that are localized on the same atom. 
The appeal of DFT+U relies on its effectiveness and low computational cost in correcting for the over-delocalization of the \textit{d} electrons in transition metal ions. 
Although the original DFT+U formulation is rigorous, in most cases $U$ and $J$ are  treated as empirical adjustable parameters which are often obtained by fitting the band structure to available experimental values.\cite{Cococcioni_reviewLDAU_2014} This approach requires fitting information from experimental data and its predictive value for new materials and hetero-structures is limited. 
There are a few common \textit{ab-initio} methods to derive $U$ and $J$ such as the constrained LDA (cLDA)~\cite{Springer_frequency_interaction_NI_RPA_PRB1998} and linear response  approach.\cite{Gunnarsson_DFTU_Andersen_PRB1989,Cococcioni_DFTU_LinearResponse_PRB2005} These methods are, unfortunately, computationally expensive, often requiring large supercell calculations. Moreover, although the linear response method has been widely used for open-shell systems, it is not suitable for closed-shell elements such as Zn and Cd, where the localized bands are completely full and insensitive to linear perturbations.\cite{Cococcioni_reviewLDAU_2014}

Hybrid functionals are based on the idea of computing the exact exchange energy from the Kohn-Sham wavefunctions and to mix it with the (semi)local approximation of correlation energy of DFT.\cite{Kummel2008OrbitalDFT} In this respect, the method  does not suffer of any limitations in dealing with closed-shell elements and it is very successful in predicting the energy gap for semiconductors and insulators. Clearly, some degree of exact exchange is necessary for a more accurate description of the electronic structure. However, also hybrid functionals have a somewhat empirical component, since the level of mixing is not determined from first principles.\cite{HSE06}

In order to facilitate the accurate characterization of electronic properties of materials at a low computational cost, a fundamental condition for the development of effective high-throughput quantum-mechanics frameworks for accelerated materials discovery,\cite{nmatHT_editorial,curtarolo:art86,nmatHT,curtarolo:art92} some of the authors have recently introduced the ACBN0   functional,\cite{ACBN0} a pseudo-hybrid Hubbard density functional that introduces a new {\it ab-initio} approach to compute ${U}$ and ${J}$ that 
does not contain any empirical parameter.
In this work we demonstrate that, by using  ACBN0,  we can improve substantially the predictive value of DFT calculations for the physical properties of Zn and Cd oxides and chalcogenides at a fraction of the computational cost of hybrid functionals.
The paper is organized as follows: in Sec.\ \ref{secmethodology}  we briefly discuss the ACBN0   method. In Sec.\ \ref{sec:Results}, we discuss our results for the Zn and Cd chalcogenides by comparing structural, electronic, and vibrational properties with more standard approaches.

\section{Methodology \label{secmethodology}}

ACBN0   is based on the DFT+\textit{U} energy functional as formulated by Dudarev (Ref. \onlinecite{Dudarev1998}):
\begin{equation*}
E_{\textrm{DFT}+U}=E_{\textrm{DFT}} + E_{U} -E_{DC}
\end{equation*}
where $E_{\textrm{DFT}}$ is the DFT energy calculated using a LDA or GGA functional, the two parameters ${U}$ and ${J}$ have been replaced by an effective on-site Coulomb interaction $U_{\textrm{eff}}={U}-{J}$ and $E_{DC}$ takes care of the double counting terms in the energy expansion. For technical details on the ACBN0   formalism we refer the reader to Ref.\ \onlinecite{ACBN0}. 



ACBN0   resolves the ambiguities in DFT+U by computing on-the-fly the local Coulomb ($U$) and  exchange ($J$) integrals for the specific orbitals under consideration via a procedure based on an ad-hoc renormalization of the density matrix. 
In this way, the value of $U_{\textrm{eff}}$ results a functional of the electron density and depends directly on the chemical environment and the crystalline field. 

 In traditional formulation of DFT+U, ~\cite{Dudarev1998}  the correction term $U$ was based explicitly on the localization of the \textit{d} orbitals and was not used for the \textit{p} or \textit{s} orbitals which tend to be less localized compared to \textit{d} electrons. However, in the ACBN0  formulation ${U_{\textrm{eff}}}$ is computed directly on the chosen Hubbard center from the Coulomb and exchange Hartree-Fock energies of the solid  and can be evaluated for any contributing orbital. The evaluation of $U_{\textrm{eff}}$ for any orbital symmetry is not a new concept (see for instance Ref.~\onlinecite{Andriotis_U_PSS2013}) but it is particularly relevant for the systems studied in this work.



In the current implementation of ACBN0, $U_{\textrm{eff}}$ is evaluated through a self consistent procedure where we start by calculating the electronic structure for an initial guess of $U^{d}_{\textrm{eff}}$ = $U^{p}_{\textrm{eff}}$ = 0 eV and converge to $U_{\textrm{eff}}$ within 10$^{-4}$ eV. A plot of the convergence of ${U_{\textrm{eff}}}$ with the iteration steps is provided in the Supplemental Information. Tab.\ \ref{tab:ZnX-U} and Tab.\ \ref{tab:CdX-U} provide the converged $U_{\textrm{eff}}$ values for all the semiconductors studied in this work.


ACBN0   calculations have been done using the Quantum \textsc{Espresso}\cite{quantum_espresso_2009} and WanT\cite{Calzolari_PRB69_2004,want} packages with norm-conserving pseudopotentials from the pslibrary1.0 database. A kinetic energy cut-off of 350 Ry and a k-point mesh of 12$\times$12$\times$12 was used for all total-energy calculations. 
Hybrid functional DFT calculations were performed using the Vienna \textit{ab-initio} simulation package (VASP)~\cite{vasp} with the HSE06.\cite{Heyd2003} This functional is defined by replacing 25\% of the PBE exchange interaction by a screened non-local functional with an inverse screening length of 0.2/\AA. A 6$\times$6$\times$6 Monkhorst-Pack k-point mesh was used and a plane-wave cut-off of 500 eV was used for all the HSE06 calculations.

\section{Results \label{sec:Results}}

This section discusses the structural, electronic, and vibrational properties of ZnO, ZnS, ZnSe, ZnTe, CdO, CdS, CdSe, CdTe computed with the ACBN0 functional.

Under standard conditions ZnO is stable in the wurtzite (\textit{wz}) structure,~\cite{ZnX-book} CdO in the rocksalt (\textit{rs}) structure, while the rest of the compounds are stable in the zinc-blende (\textit{zb}) structure.
Due to the importance of these materials in semiconductor nano-structures such as heterojunctions and quantum dots that are usually grown epitaxially with far-from-the-equilibrium techniques, we study all the three competing phases for all the chemical compositions. These different structures can be stabilized by alloying and epitaxial strain. 
Results not reported in the manuscript can be found in the Supplemental Information.

\subsection{Structural properties}

\begin{figure}
\includegraphics[scale=0.45]{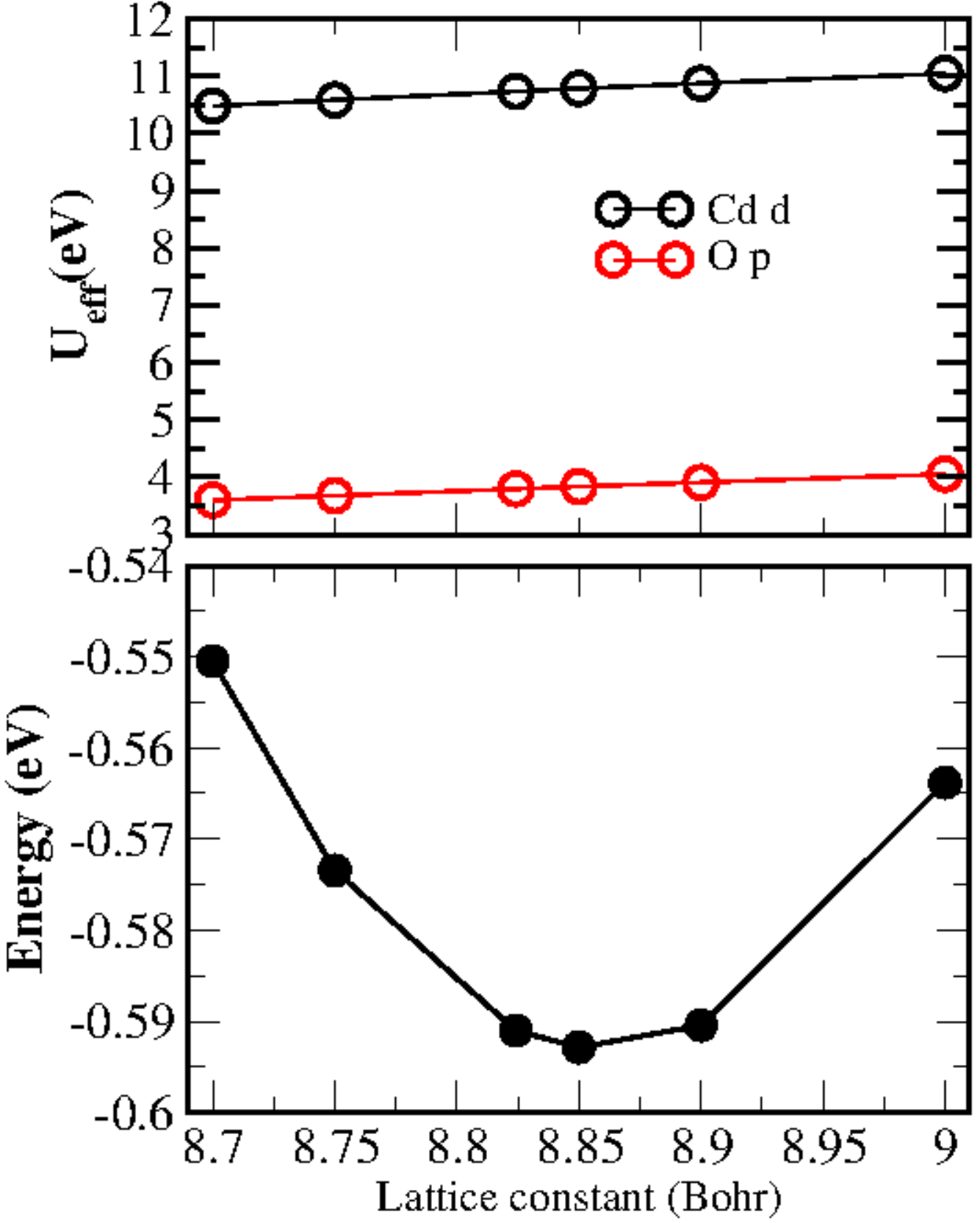}
\caption{\label{fig:CdO-Ueff} (Color online) {Lower panel:} Energy-volume curve of CdO in the \textit{rs} phase as calculated by the ACBN0 functional. {Upper panel:} Variation of the converged $U_\textrm{eff}$ of Cd and O in CdO as a function of lattice constant.}
\end{figure}

The lattice constants ($a_0$) and bulk moduli ($B$) for each of the II-VI semiconductors in the three bulk phases
are obtained by fitting the total energy as a function of volume to the Murnaghan's equation of state (eos). 
As discussed in Sec.\ \ref{secmethodology}, ${U_{\textrm{eff}}}$ is a functional of the electron density and is thus dependent on the geometry of the ground state. 
For a small volume change around equilibrium, the calculated ${U_{\textrm{eff}}}$  varies linearly with the volume.  As an example, In the top of Fig.~\ref{fig:CdO-Ueff} we have plotted the variation of $U_{\textrm{eff}}$ for CdO in the \textit{rs} phase where we observe the linear scaling for both the Cd's and the O's $U_{\textrm{eff}}$, most likely related to bond length changes that affect the wave-functions overlap and the electron localization.

\begin{table}[htbp]
\caption{\label{tab:ZnX-U}Converged  values of ${U_\textrm{eff}}$ (in eV) for the Zn \textit{3d} and the anion \textit{p} states. The calculations were performed at the  equilibrium lattice constants in Table \ref{tab:lattice-constant}.}
\begin{tabular}{lc|c|c}
\hline
&Phase &  Zn\textit{3d} & Anion \textit{p} \\ 
\hline
ZnO & wz &  13.19   & 5.57  \\
    & zb &  13.24&  6.02\\
    & rs &  13.52&  6.05\\
\hline
ZnS & wz& 13.48    &3.50   \\
    & zb & 13.21   & 3.42 \\
    & rs & 14.54   &  3.76\\
\hline
ZnSe & wz& 15.00    &2.55   \\
    & zb &14.60    &2.52  \\
    & rs &  15.68&  2.77\\
    \hline
ZnTe & wz & 16.00 & 2.40 \\
     & zb & 16.89 & 2.43 \\
     & rs & 17.08 & 2.66 \\
\hline

 \end{tabular}
\end{table}

\begin{table}
\caption{\label{tab:CdX-U}Converged values of  ${U_\textrm{eff}}$ (in eV) for the Cd \textit{4d} and the anion \textit{p} states. The values are for the equilibrium lattice constants in Table \ref{tab:lattice-constant}.}
\begin{tabular}{lc|c|c}
\hline
Phase &  & Cd \textit{4d} & Anion \textit{p} \\ 
\hline

CdO & wz &  10.69  & 4.50  \\
    & zb &  10.49 & 4.03 \\
    & rs &  10.73&  3.92\\
\hline
CdS & wz&  11.07   & 3.69   \\
    & zb & 10.89   &3.64  \\
    & rs & 11.77&  3.94\\
\hline
CdSe & wz& 11.58    & 2.68   \\
    & zb & 11.69  & 2.70  \\
    & rs & 12.51 & 3.05 \\
\hline
CdTe & wz& 12.95    & 2.57   \\
    & zb & 13.00   & 2.55  \\
    & rs &  13.70& 2.46 \\
\hline

\end{tabular}
\end{table}

In Tables.~\ref{tab:lattice-constant} and \ref{tab:bulk-modulii}, we have listed all the values of the computed lattice constants and bulk moduli of all the II-VI semiconductors in the three phases using PBE, HSE and ACBN0   (and SIRC, whenever available) with references to existing experimental values.
Experimentally, the lattice constants follow the trend of ZnO $<$ ZnS $<$ZnSe $<$ ZnTe and similarly for the Cd\textit{X} series. This trend of increasing lattice constant is described in the PBE, HSE and ACBN0   functionals. The experimental bulk moduli shows a reverse trend \textit{i.e.}, the bulk modulus decreases as we go from O to S to Se to Te.

\begin{figure}
\includegraphics[scale=0.40]{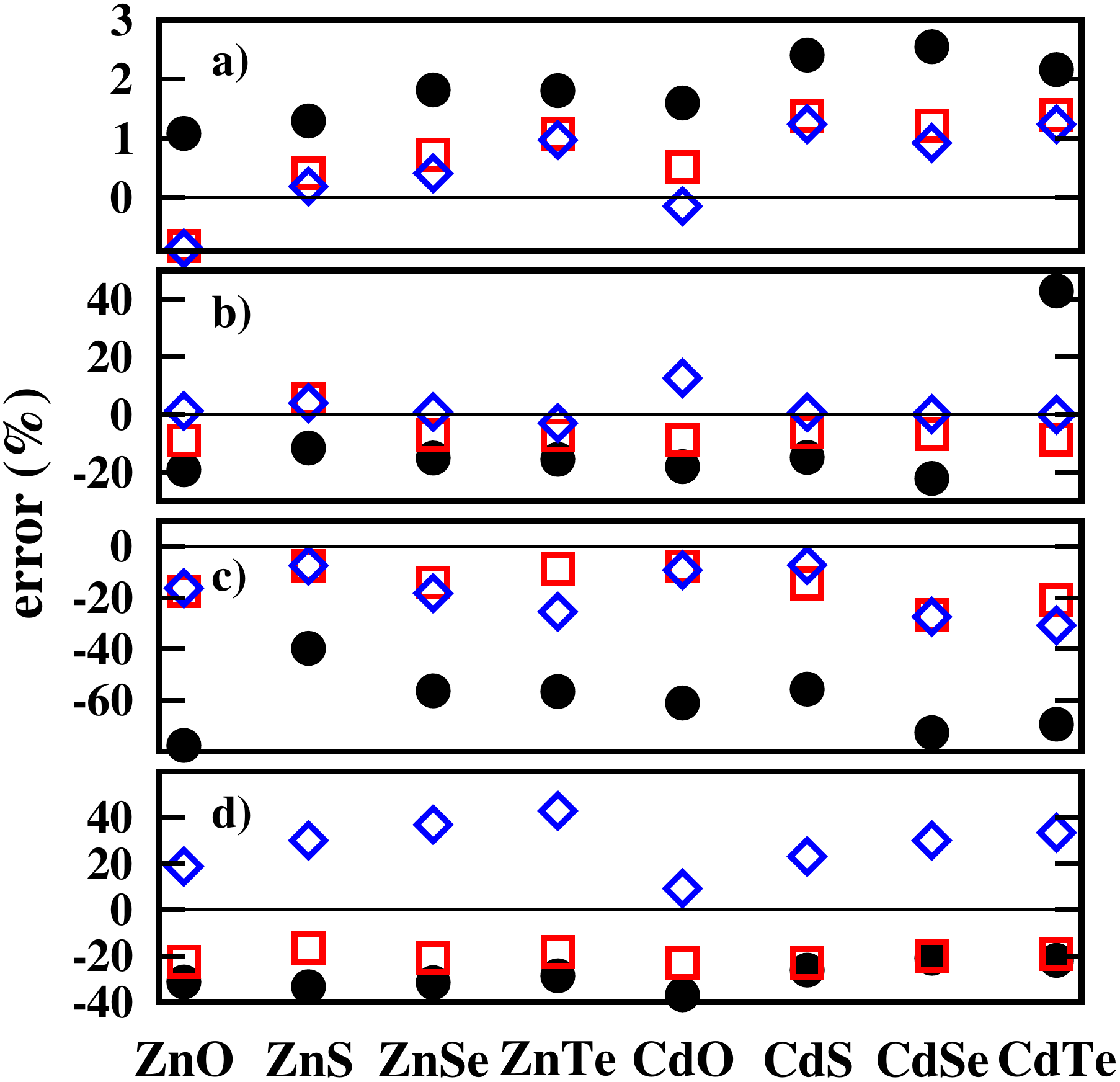}
\caption{\label{fig:errors}(Color online) Comparison of percentage relative error in the predictions of (a) lattice constants (b) bulk-moduli (c) band-gap (d) energies of the \textit{d} band calculated using PBE, HSE, and ACBN0 functionals. All the absolute values are listed in Tables~[\ref{tab:lattice-constant}-\ref{tab:band-gap}].~The errors are calculated relative to the experimental values in Tables~[\ref{tab:lattice-constant}-\ref{tab:band-gap}].~In cases where there is more than one experimental value, we take the average of the minimum and the maximum value as the reference experimental value. Comparison is made only for the stable phases of the II-VI compounds. Black circles are the PBE values, open red squares are the HSE values and open blue diamonds are the ACBN0   values. The experimental values is referenced by the black solid line at 0.}
\end{figure}

In Figures~\ref{fig:errors}(a) and \ref{fig:errors}(b), we have plotted the relative percentage error of the three functionals PBE, HSE, and ACBN0   in predicting the lattice constants and bulk moduli with respect to the measured values. In the plot, we have included only the stable phases for each of the eight semiconductors since we did not have experimental data on the non-equilibrium phases. The values for the other three phases for the II-VI semiconductors are listed in Table~\ref{tab:lattice-constant}. Except for CdO and ZnO, all the other semiconductors have a stable \textit{zb} structure. For ZnO, we have listed the error for one of the lattice parameters, $a$, while for CdO we have used the \textit{rs} phase. 

The lattice constants and bulk modulii are more accurately described  by HSE and ACBN0   compared to PBE functional calculations. The PBE overestimates the lattice constants significantly (up to 2\%) as seen in Fig.~\ref{fig:errors}(a) and severely underestimates the bulk-moduli (See Fig.~\ref{fig:errors}(b)); in some cases the error is more than 50\% compared to the experimental value. 
The HSE functional shows a better improvement in the prediction of lattice constants reducing the error to less than 1\%. The lattice constants predicted by the ACBN0   follows the HSE closely with an error less than 1\%. Note that the ACBN0   preserves the accuracy of the HSE calculations with 8-10 times speed up in terms of calculation time. 

The bulk modulus predicted by ACBN0  shows a tremendous improvement over both the HSE and the PBE functional. In CdS, CdSe and CdTe, the agreement with experimental values is exceptional with an error less than (0.5\%). Note that the PBE error in the bulk moduli are greater than 20\%. These results indicates that using ACBN0  for the calculations of phonon spectra or any lattice dynamical property will yield much improved results, as already observed in the case of ZnO,\cite{Nardelli_Scientific_Reports_2013} where the values of the Zn and O $U_{\textrm{eff}}$'s, originally fitted to reproduce the experimental band gap and position of the \textit{d} bands, are in remarkable agreement with the ones predicted by ACBN0. Incidentally, $U_{\textrm{eff}}$ values fitted to HSE calculations have been proposed in the past for ZnO and ZnS.\cite{Andriotis_U_PSS2013,Picozzi_spin_PRB2012} While for ZnO the values are again close to our predictions, the values for ZnS are very much different. See Sec. \ref{phonon} for a comprehensive discussion of the lattice dynamical properties of these systems. 

For all other phases, the ACBN0   functional reproduces the HSE values closely as seen in Table~\ref{tab:lattice-constant} and ~\ref{tab:bulk-modulii} at a much lower computational cost. Overall the the structural properties of the II-VI semiconductors computed with the ACBN0 functional show better agreement with respect to experimental results.

\begin{table*}
\caption{Comparison of the lattice constants for the different compounds calculated using ACBN0   and other functionals. All the PBE and HSE values are computed in this work. For \textit{wz}, the first value is the in-plane lattice parameter $a$ and the second value is the out-of-plane lattice parameter $c$. All the SIRC values reported here are taken from Ref.~\onlinecite{Vogel_self_interaction_pseudopotentials_II_IV_PRB1996}}.
\begin{ruledtabular}
\begin{tabular}{lcccccc}

  System & Phase & PBE &  HSE & SIRC & ACBN0  & Experiment   \\

\hline
 ZnO    & wz & 3.283, 5.309  &  3.260, 5.221 & 3.290, 5.29 &  3.270, 5.164 & 3.258, 5.220~\cite{Landolt-Bornstein} \\ 
        & zb & 4.670         & 4.582         &             & 4.580         &4.620~\cite{Ashrafi_ZnOZB_JAP2007,Uddin_ZnOphase_PRB2006}  \\
        & rs & 4.370         & 4.278         &             & 4.289        & 4.272~\cite{Uddin_ZnOphase_PRB2006,fan_ZnObulk_APL08} \\
        \hline
 ZnS     & wz &3.880,6.300  & 3.850,6.271    & 3.830,6.280  & 3.851, 6.278 & 3.811, 6.234~\cite{grunwald_CdS_JCP2012}\\
         & zb & 5.489       & 5.432          &5.421         &  5.437       & 5.410~\cite{Landolt-Bornstein}          \\
         & rs & 5.143       & 5.077          &          & 5.080        &   5.060~\cite{jaffe_ZnS_PRB1993,grunwald_CdS_JCP2012}                         \\
         \hline 
 ZnSe    &wz  & 4.043,6.70   & 4.030,6.620     &            & 4.020, 6.630   & 3.980,6.530~\cite{Landolt-Bornstein,Vogel_self_interaction_pseudopotentials_II_IV_PRB1996}\\     
         &zb  & 5.771       &   5.708        &            &  5.696       & 5.667 ~\cite{Landolt-Bornstein,Heyd_bandgaps_JCP2005}\\
         &rs & 5.401   &  5.330               &            &  5.321       &    \\
         \hline
 ZnTe    & wz & 4.366,7.176    &  4.35,7.132         & & 4.340, 7.140     &  4.320,7.100~\cite{Lakshmi-ZnTe-2003}                             \\
         & zb & 6.199          & 6.154       &          &   6.148        &  6.089~\cite{Vogel_self_interaction_pseudopotentials_II_IV_PRB1996,Heyd_bandgaps_JCP2005}                            \\
         & rs & 5.788          & 5.743       &         &   5.732         &                           \\
         \hline   
 CdO    & wz & 3.700,5.850   & 3.650,5.800   &     &  3.590,5.750          &                           \\
         & zb & 5.172      & 5.103           &     &  5.054              &                           \\
         & rs & 4.771      & 4.722           & 4.781 & 4.689               & 4.696~\cite{Schelfe_CdO_PRB2006,Burbano_CdO_ACS2011,Mudd_CdOldau_PRB2014}                          \\
         \hline
 CdS    & wz & 4.206,6.85  & 4.223,6.951     & 4.154,6.762               & 4.183,6.682 &  4.135,6.701~\cite{Liwei-EPL-CdS-2014,grunwald_CdS_JCP2012} \\ 
        & zb &  5.963     &  5.901          &5.851  &   5.892            & 5.820~\cite{Hinuma-CdS-HSE-2013, Landolt-Bornstein,grunwald_CdS_JCP2012}                            \\
         & rs & 5.890      &  5.466       & & 5.459                     &      \\
         \hline
 CdSe    & wz & 4.39,7.1  &4.350,7.112     & 4.292,7.021 & 4.33,7.072   & 4.310,7.010~\cite{Vogel_self_interaction_pseudopotentials_II_IV_PRB1996}                             \\
         & zb & 6.239      &  6.158       & 6.071 &   6.142            & 6.084~\cite{Vogel_self_interaction_pseudopotentials_II_IV_PRB1996}                          \\
         & rs & 5.800      & 5.697          & &  5.680                  &               \\
         \hline        
CdTe   & wz & 4.550, 7.451     & 4.580,7.460            &  & 4.600, 7.498 &                              \\
         & zb & 6.621      &  6.571     & 6.401 &    6.560             &  6.480~\cite{Heyd_bandgaps_JCP2005}                    \\
         & rs & 6.130      &  6.092       &  & 6.076      &                                    
\end{tabular}
\end{ruledtabular}
\label{tab:lattice-constant}
\end{table*}

\begin{table*}
\caption{\label{tab:bulk-modulii}Comparison of the bulk moduli $B$ (GPa) of different compounds using three different functionals. All the PBE and HSE values are calculated in this work.}
\begin{ruledtabular}
\begin{tabular}{lcccccc}
System & Phase & PBE & HSE & SIRC& ACBN0   &Experiment  \\
\hline
ZnO    & wz    & 127 & 143  & 159 & 157 & 136-183~\cite{Uddin_ZnOphase_PRB2006,Landolt-Bornstein,Vogel_self_interaction_pseudopotentials_II_IV_PRB1996} \\
       & zb    &123  & 143  & & 162  &                 \\
       & rs    & 156 & 188  &   & 205  & 177-228~\cite{Uddin_ZnOphase_PRB2006}  \\    
         \hline
ZnS    & wz & 60      & 74  &81  & 79          &76~\cite{kisi-ZnS-1989-ActaCrys}  \\       
         & zb & 68    & 74.4  &  81 & 80 & 76.9~\cite{Landolt-Bornstein} \\  
         & rs & 84    &  95.4  &   &  101.6   &  103.6~\cite{jaffe_ZnS_PRB1993}                       \\
         \hline
 ZnSe    & wz & 57   &  62.69  &  & 64     &                              \\
         & zb &  55.8  & 61    &   &  66.3  &  65.7~\cite{Heyd_bandgaps_JCP2005}                              \\
         & rs &  70       & 78    &      & 66.8    &                            \\
         \hline
 ZnTe    & wz & 45      & 58.46          &  &56    &                          \\
         & zb & 43   &  47.19   &  &  49.4    & 50.9~\cite{Vogel_self_interaction_pseudopotentials_II_IV_PRB1996,Heyd_bandgaps_JCP2005}        \\
         & rs & 53.3   &  58.82     &  &   60     &                              \\
         \hline   
 CdO    & wz & 92.7      & 104          &  & 114    &                              \\
         & zb & 91       & 102      &  & 124       &                              \\
         & rs &  119     &   137      & 152 &    170    & 147~\cite{Liu-APL-CdO-2006}\\       
   \hline
 CdS    & wz &  54     &  57           &74  & 53    &  61~\cite{Vogel_self_interaction_pseudopotentials_II_IV_PRB1996}, 65~\cite{grunwald_CdS_JCP2012}                            \\
        & zb & 53.5    &  58.94        & 70 &  63.3  & 55~\cite{grunwald_CdS_JCP2012} \\ 
         & rs & 63.3    & 78.6        &  & 83.2    &        \\
         \hline        
 CdSe    & wz &  48     &    50.80       & 62 &  52   & 55                             \\
         & zb & 41.5      & 49.7   & 66  &  53.4       & 45.1~\cite{grunwald_CdS_JCP2012}                            \\
         & rs &  53     &  67.4    &  & 64            &                              \\
         \hline        
CdTe   & wz &  55     &  60         &  &  57.4                                                  &                              \\
         & zb & 57      &   38.4        & 52 & 42     & 42~\cite{Landolt-Bornstein}                              \\
         & rs &  45     &   51.86       & & 53.1                                               &                                    

\end{tabular}
\end{ruledtabular}
\end{table*}

\subsection{Electronic properties}

The band-structures of the eight semiconductors in all the three different crystal phases were computed to assess the veracity of the ACBN0   functional in the prediction of band gaps. For comparison, we computed the band-structures within the PBE and the HSE functionals. All the band structures were calculated at the theoretical equilibrium volume optimized within each functional. 
In the \textit{wz} and \textit{zb} compounds, the band-gap is direct while in the \textit{rs} compounds the gap is indirect between the $L$ and $\Gamma$ point. A summary of the values of the energy-gaps ($E_g$) for the different compounds is presented in Table~\ref{tab:band-gap} as calculated by the ACBN0, PBE and HSE functionals along with references to the SIRC and experimental values whenever available. 

\begin{table*}
\caption{\label{tab:band-gap} Comparison of the energy-gaps ($E_g$) in eV for the different compounds using three different functionals. All the PBE and HSE values are calculated in this work. In the \textit{rs} phase, two values are listed. The first one is the direct band gap ($\Gamma$-$\Gamma$) while the second one in parenthesis is the indirect band gap between $L$ and $\Gamma$ point. For the \textit{zb} and \textit{wz} phases, the gap is direct from $\Gamma$-$\Gamma$. The experiment values are reported whenever available.}
\begin{ruledtabular}
\begin{tabular}{lcccccc}
System & Phase & PBE &  HSE & SIRC& ACBN0   & Experiment \\
\hline
ZnO     & wz  & 0.85& 2.90 & 3.4  &2.91  & 3.2 ~\cite{Landolt-Bornstein}, 3.4~\cite{Uddin_ZnOphase_PRB2006}\\
        & zb &  0.59     & 2.70          & & 2.74  &    3.27~\cite{Ashrafi_ZnOZB_JAP2007}                          \\
        & rs & 2.17 (0.89)       &  4.35(2.90)     &  &       4.17(3.09)   &   4.5(2.45)~\cite{Kuang-ZnO-HSE-phase-2014}                           \\
         \hline
 ZnS    & wz &  2.10 &  3.42, 3.34~\cite{yadav2012strain}  &3.6  & 3.31 &   3.86~\cite{kisi-ZnS-1989-ActaCrys}, 3.91~\cite{kasap2007springer}                       \\
        & zb & 2.23    & 3.49~\cite{peverati_M112012}        &3.6  & 3.42    &  3.7~\cite{Springer_frequency_interaction_NI_RPA_PRB1998}, 3.66~\cite{Heyd_bandgaps_JCP2005}                            \\
        & rs & 2.1(0.83)      &  3.7(1.31)        & &   3.52(1.28)                                              &                             \\
         \hline
 
 ZnSe    & wz & 1.27      & 2.46          &  &  1.90                                                  &2.87~\cite{}                              \\
         & zb & 1.18      & 2.32~\cite{Heyd_bandgaps_JCP2005},2.42~\cite{peverati_M112012}          &2.1  &2.041                                                    & 2.70~\cite{Heyd_bandgaps_JCP2005}                             \\
         & rs & 1.4($<$0)  & 2.60($<$0)   &  &  2.02($<$0)      &                              \\
         \hline
 ZnTe    & wz & 1.17      &  2.22         &  & 1.82                &                              \\
         & zb & 1.043 & 2.36~\cite{peverati_M112012},2.19~\cite{Heyd_bandgaps_JCP2005}  &1.4  & 1.79 & 2.38~\cite{Heyd_bandgaps_JCP2005}                             \\
         & rs & 0.6   & 1.4        &  &  1.2        &                              \\
         \hline   
 CdO    & wz &   $<$0   & 0.75, 1.13~\cite{yan2012strain}      &  & 1.23                                               &  0.91~\cite{GaUJ}                              \\
         & zb &  $<0$    & 1.04          & & 1.35        &                              \\
         & rs & 0.85($<0$) & 2.01(0.89) &  & 1.98(0.70) & 2.18-2.23(0.9-1.08)~\cite{Mudd_CdOldau_PRB2014,Faragani-CdO-APL-2013}                              \\
         \hline
 CdS    & wz &  1.36     & 2.15, 2.09~\cite{Liwei-EPL-CdS-2014}         & 2.5 & 2.4    & 2.5~\cite{rossow-1993-Thinfilms}                             \\
         & zb & 1.11 & 2.13      & 2.4 &2.3                                             & 2.55~\cite{Heyd_bandgaps_JCP2005}, 2.4~\cite{Vogel_self_interaction_pseudopotentials_II_IV_PRB1996}                             \\
         & rs & 1.71($<$0)       &  2.9(1.4)      &  &     2.95(1.27)               & 1.5~\cite{Heyd_bandgaps_JCP2005}                              \\
         \hline        
 CdSe    & wz & 0.73      &  1.77         &1.3 & 1.61    & 1.8~\cite{Vogel_self_interaction_pseudopotentials_II_IV_PRB1996}                             \\
         & zb & 0.52      &1.39       &1.4  & 1.38   & 1.90~\cite{Heyd_bandgaps_JCP2005},1.82~\cite{Vogel_self_interaction_pseudopotentials_II_IV_PRB1996}                             \\
         & rs & $<$0      & 0.69($<$0)          &  &   1.4(0.3)    &                             \\
         \hline        
CdTe   & wz &  0.64     & 1.6          &  &  1.33   & 1.8                             \\
         & zb & 0.58 & 1.51, 1.52~\cite{xiao-2011-PCL-bandgap},1.67~\cite{peverati_M112012}        & 0.8 &  1.43                                                 & 1.92~\cite{Heyd_bandgaps_JCP2005},1.61~\cite{xiao-2011-PCL-bandgap}                           \\
         & rs & ($<$0)    & 1.4($<$0)          &  & 1.3($<$0)                                                  &                                    

\
\end{tabular}
\end{ruledtabular}
\end{table*}

Experimentally, the band gap in both the Zn\textit{X} and the Cd\textit{X} series decreases with increasing lattice constants \textit{i.e.}, the band gap decreases as we proceed from S to Se to Te. Both ZnO and CdO show the 
anomaly of a lower band-gap compared to the ZnS and CdS respectively, in spite of a lower lattice constant. This trend of decreasing band gap is described by all the functionals: PBE, HSE and ACBN0. 
The absolute value of the band-gap, however, is severely underestimated by PBE: in the oxides, especially, we see an error bar of more than 60\% (See Fig.\ref{fig:errors} (c)). The ACBN0   functional reduces this error to less than 20\%. The band-gaps predicted by ACBN0   for the eight semiconductors studied here lie within 0.5 eV of the experimental value. 

Before discussing the similarities of the HSE and the ACBN0   functionals, we first look at the general features of the calculated band-structures. Representative band-structures for the eight semiconductors in their stable phases calculated with the ACBN0   and the PBE functional are shown in Figures~\ref{fig:ZnX-band} and \ref{fig:CdX-band} (band structure of all the other phases are compiled in the Supplementary Information). 
Within PBE, (black lines), the overall band profiles are very similar in both the Zn\textit{X} and Cd\textit{X} series. The lowest energy manifolds (below -10 eV) are rather narrow and entirely of \textit{s} character. 
At higher energy we found the bands derived from the \textit{d} orbitals of Zn and Cd that are just below (or little hybridized) with the oxygen and chalcogen \textit{p} states forming the manifolds at the top of the valence band. The conduction band is dominated by the cation \textit{s} states. As we proceed from S to Te, the chalcogen \textit{p} states shift up in energy and this results in the disentangling of the \textit{d} bands. 
Compared to ZnS, in CdS the Cd \textit{d} bands are already well-separated from the S \textit{p} bands. 
As seen from the PBE band-structures, in both ZnO and CdO the \textit{d} states are more dispersed compared to the chalcogenides. The presence of a spurious \textit{p}-\textit{d} hybridization in ZnO is a macroscopic manifestation of the inadequacy of traditional (semi)local functionals to predict correctly the electronic properties of materials where strong electron localization is present and  results in the strong underestimation of the band-gap: 0.85 eV versus an experimental value of 3.4 eV. Finally, CdO is predicted to be a semi-metal by PBE.
As it is clearly evident from the figures, the ACBN0   bands (red lines) maintain their overall character as discussed above but display a much wider separation between valence and conduction bands. The relative position of the \textit{d} bands is shifted to lower energies, thus promoting the disentanglement of the Zn $d$ and O $p$ manifolds. For instance, the  ACBN0   converged values for ZnO and CdO energy gaps are 2.91 eV and 0.70 eV, respectively, in better agreement with the experimental values of 3.4 eV and 0.9 eV. At the same time, the Zn $d$ bands of ZnO are shifted by  approximately 4 eV with respect to the O $p$ bands, removing the spurious hybridization present in PBE. 


Note that the ACBN0   follows the HSE functional in predicting the band-gaps. Both the ACBN0   and the hybrid functionals reduce the static correlation energy associated with the localized orbitals and hence provide a similar correction. Indeed, Ivady \textit{et.al
} (Ref. \onlinecite{Ivady2014Theoretical_unification}) have recently shown that the hybrid exchange correlation potential can be rewritten mathematically as an on-site Hubbard potential for a system of localized orbitals, thus providing a formal justification for our observations. 

\begin{table}
\caption{\label{tab:denergy} Average band energies (E$_d$) in eV for the eight compounds in their stable phases as calculated by PBE, ACBN0 and HSE functionals. The experimental values are taken from Ref.~\onlinecite{Vogel_self_interaction_pseudopotentials_II_IV_PRB1996}. For CdO, the value is taken from Ref.~\onlinecite{Mudd_CdOldau_PRB2014}}
\begin{tabular}{lcccc}
\hline
\hline
Compound & PBE & HSE & ACBN0 & Experiment \\
\hline
ZnO & -5.5  & -6.2  & -9.3 & -7.8 \\
ZnS &  -6.0   & -7.5  & -11.7 & -9.0     \\
ZnSe & -6.5 & -7.5  &  -12.9  & -9.4      \\
ZnTe & -7.0 & -8.0 & -13.8 & -9.8\\
CdO  &  -5.5   & -6.8       & -9.5     & -8.7     \\
CdS &    -7.0  &   -7.4    &  -12.0    &  -9.6     \\
CdSe &  -7.8   &   -8.0    &   -13.0   &  -10.0     \\
CdTe &  -8.3  &     -8.4  &   -14.0   &  -10.5      \\

\hline

\end{tabular}
\end{table}
In Fig. \ref{fig:errors}(d) we have plotted the error bar in the position of the $d$ manifold of  Zn and Cd in PBE, HSE and ACBN0. We clearly observe the limits of ACBN0   and HSE  in predicting the energy of the \textit{d} bands relative to the valence band maximum. In PBE and HSE, we see a negative error of around 40\% which indicates that the position of the \textit{d} bands are higher in energy compared to the experimental values while in ACBN0, the bands are pushed much lower, closer to the experimental position. 
The reason for the overestimation of the \textit{d} band energy in HSE calculations is due to the incomplete treatment of correlation energy of the localized \textit{d} electrons. In ACBN0, downward shifting of the 3${d}$ or 4${d}$ bands increase monotonically with increasing values of $U_{\textrm{eff}}$. The rigid shift of the bands arises from a singularity due to the filled character of the $d^{10}$ bands and it is implicit in the definition of the Hubbard correction. An extensive discussion  for the case of  ZnO is provided in Ref. \onlinecite{ACBN0}. 

\begin{figure*}
\caption{\label{fig:ZnX-band} (Color online) Band structure of Zn\textit{X} (\textit{X}=O,S,Se,Te) within the ACBN0 (Red lines) method for the stable \textit{wz} and \textit{zb} phase. The PBE (Black lines) band structure is also shown for comparison. Brillouin zone integration follows the AFLOW standard as discussed in Ref.~\onlinecite{WahyuBZ}}

\begin{tabular}{cccc}
 \includegraphics[scale=0.25]{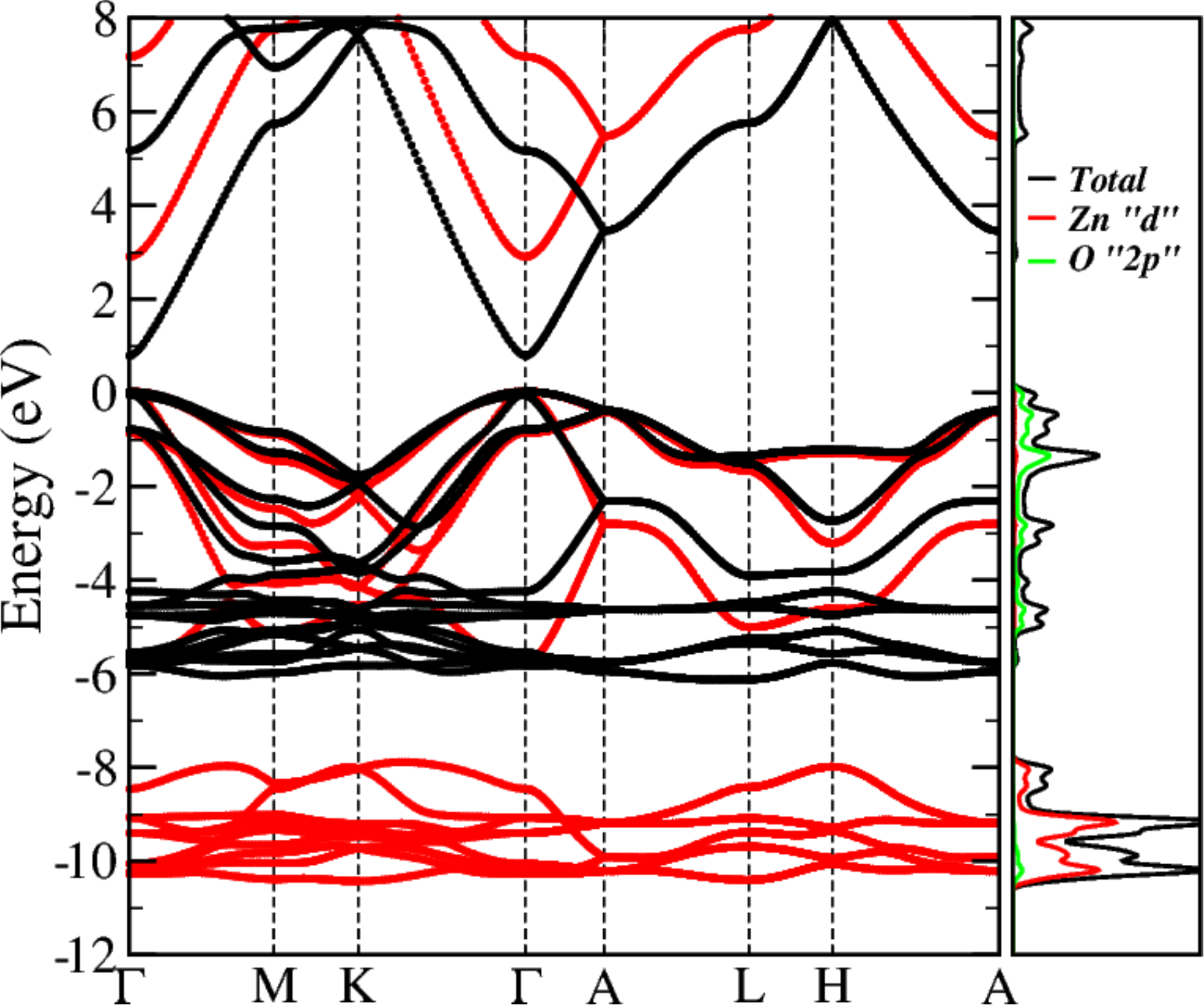}& \includegraphics[scale=0.25]{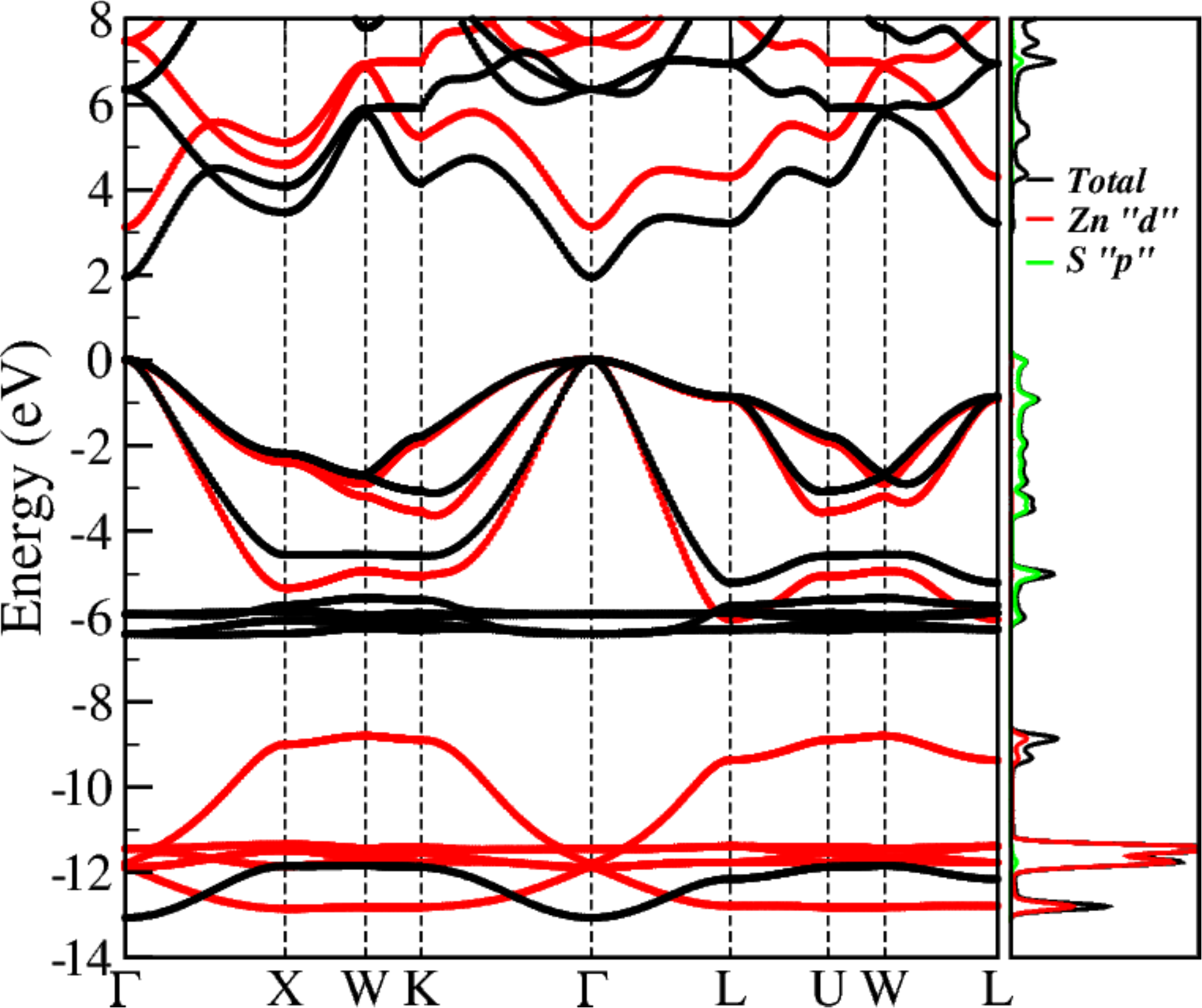} & \includegraphics[scale=0.25]{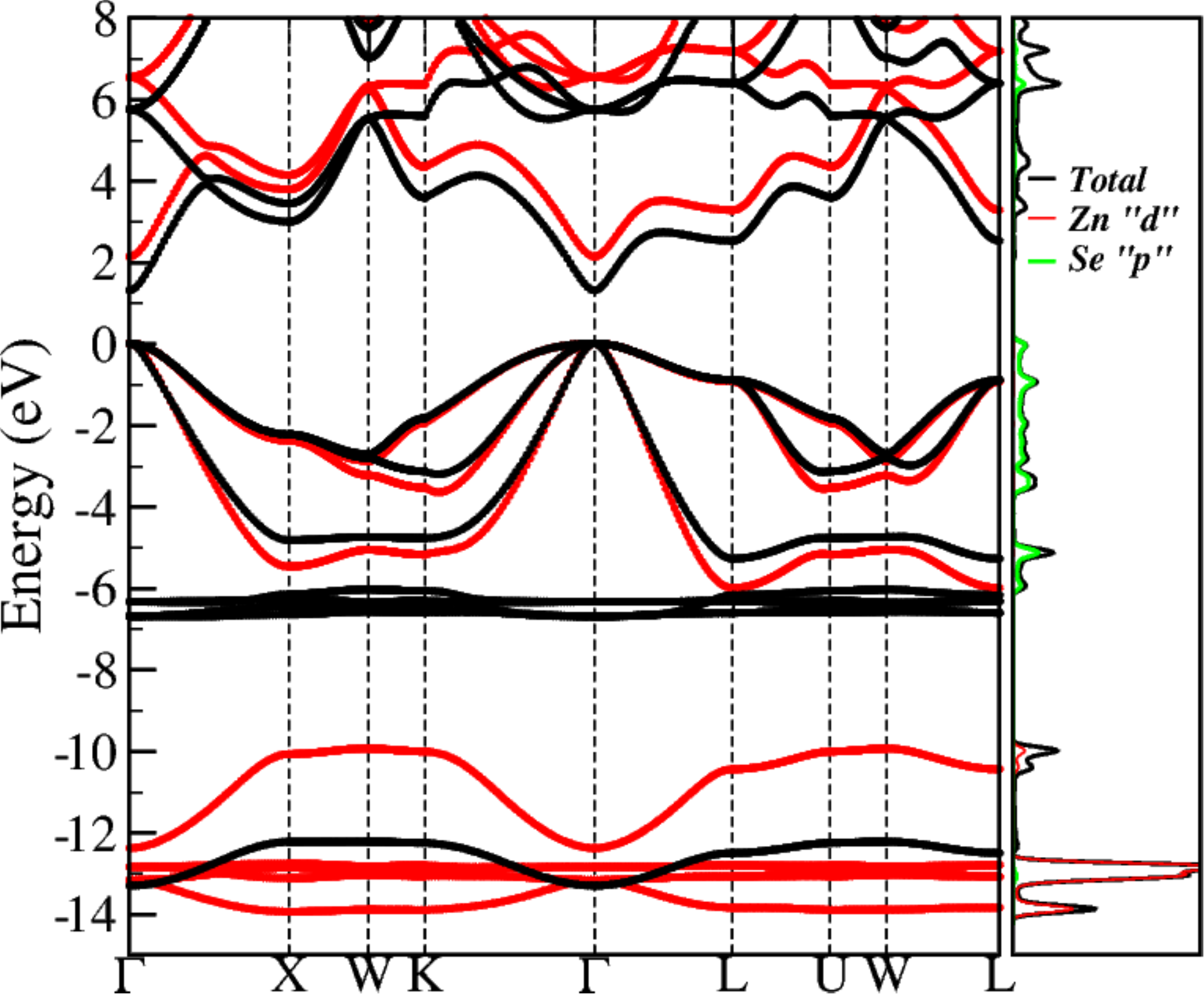}& \includegraphics[scale=0.25]{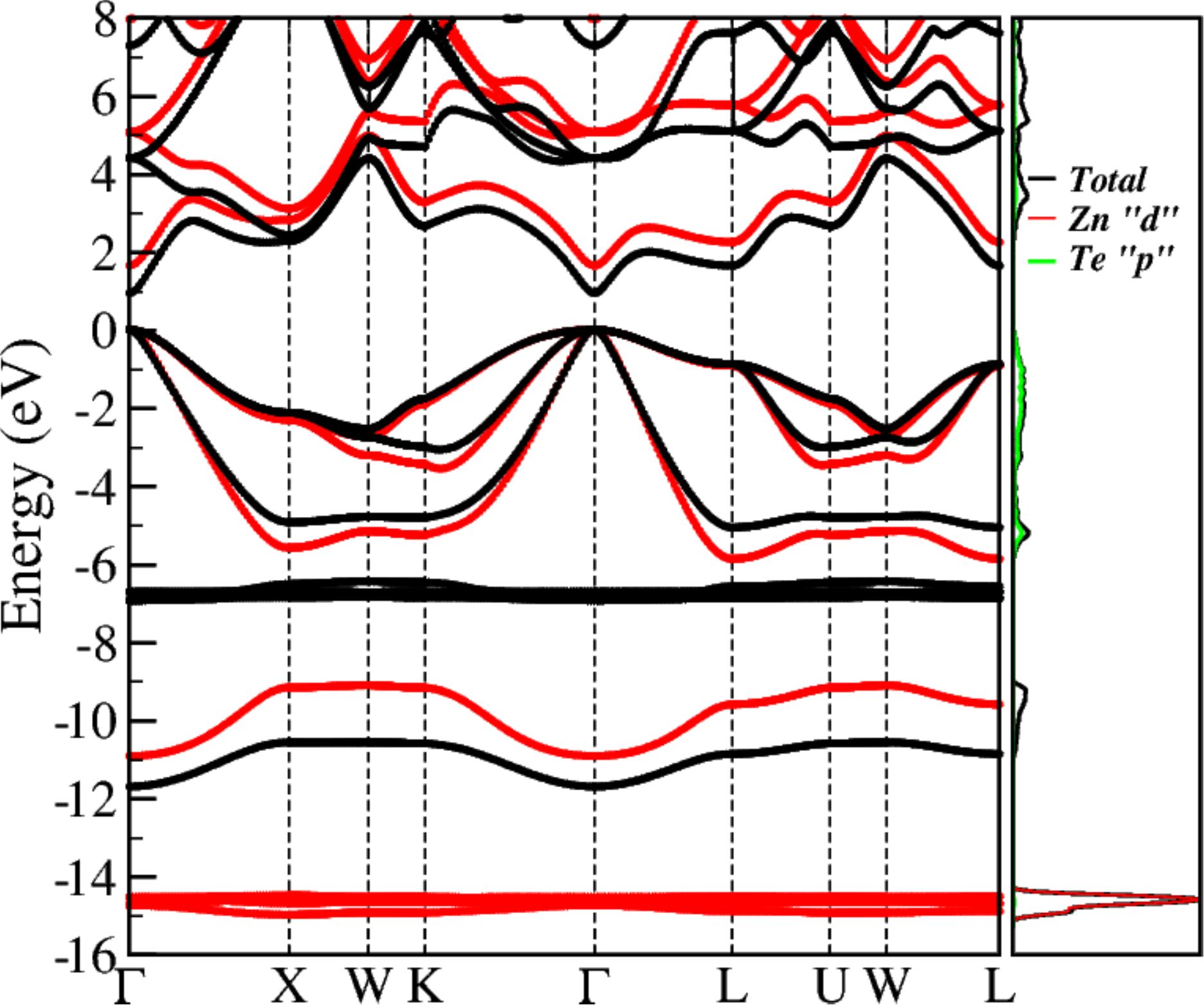} \\
 ZnO&ZnS & ZnSe &ZnTe \\
\end{tabular}
\end{figure*}

\begin{figure*}
\caption{\label{fig:CdX-band}(Color online) Band structure of Cd\textit{X} (\textit{X}=O,S,Se,Te) within the ACBN0 (Red lines) method for the stable \textit{rs} and \textit{zb} phase. The PBE (black lines) band structure is also shown for comparison. Brillouin zone integration follows the AFLOW standard as discussed in Ref.~\onlinecite{WahyuBZ}}
\begin{tabular}{cccc}
\includegraphics[scale=0.25]{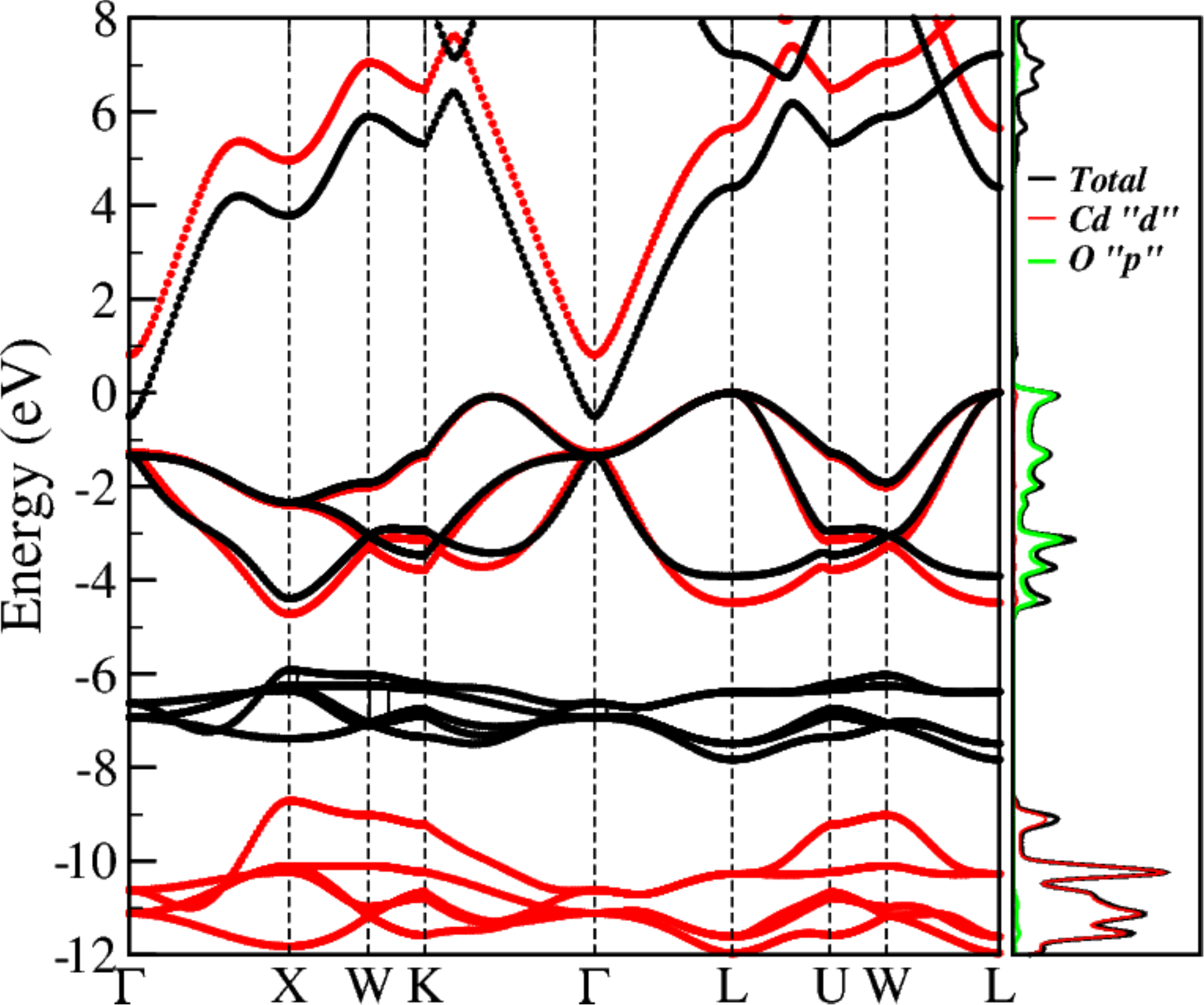} & \includegraphics[scale=0.25]{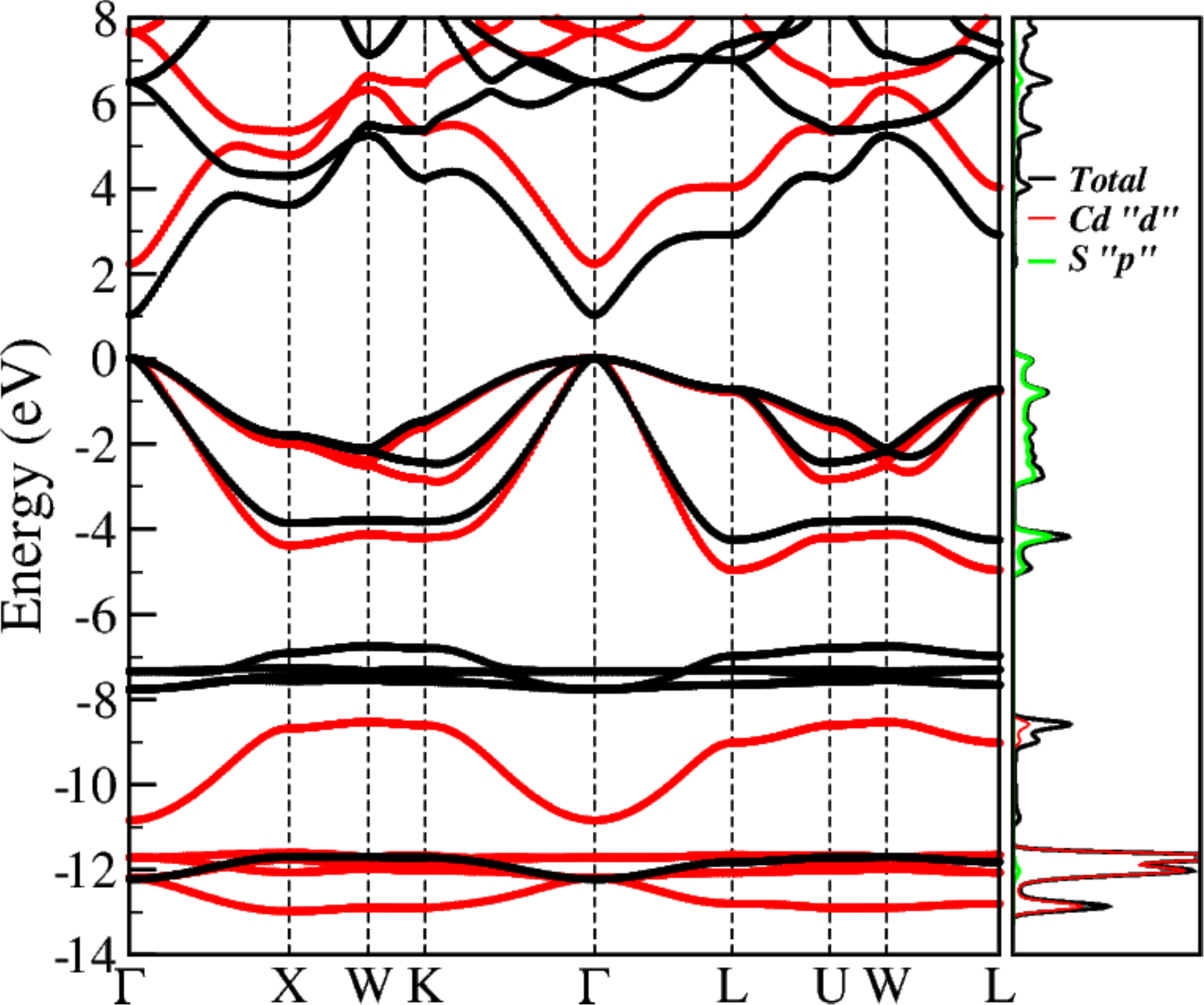} & \includegraphics[scale=0.25]{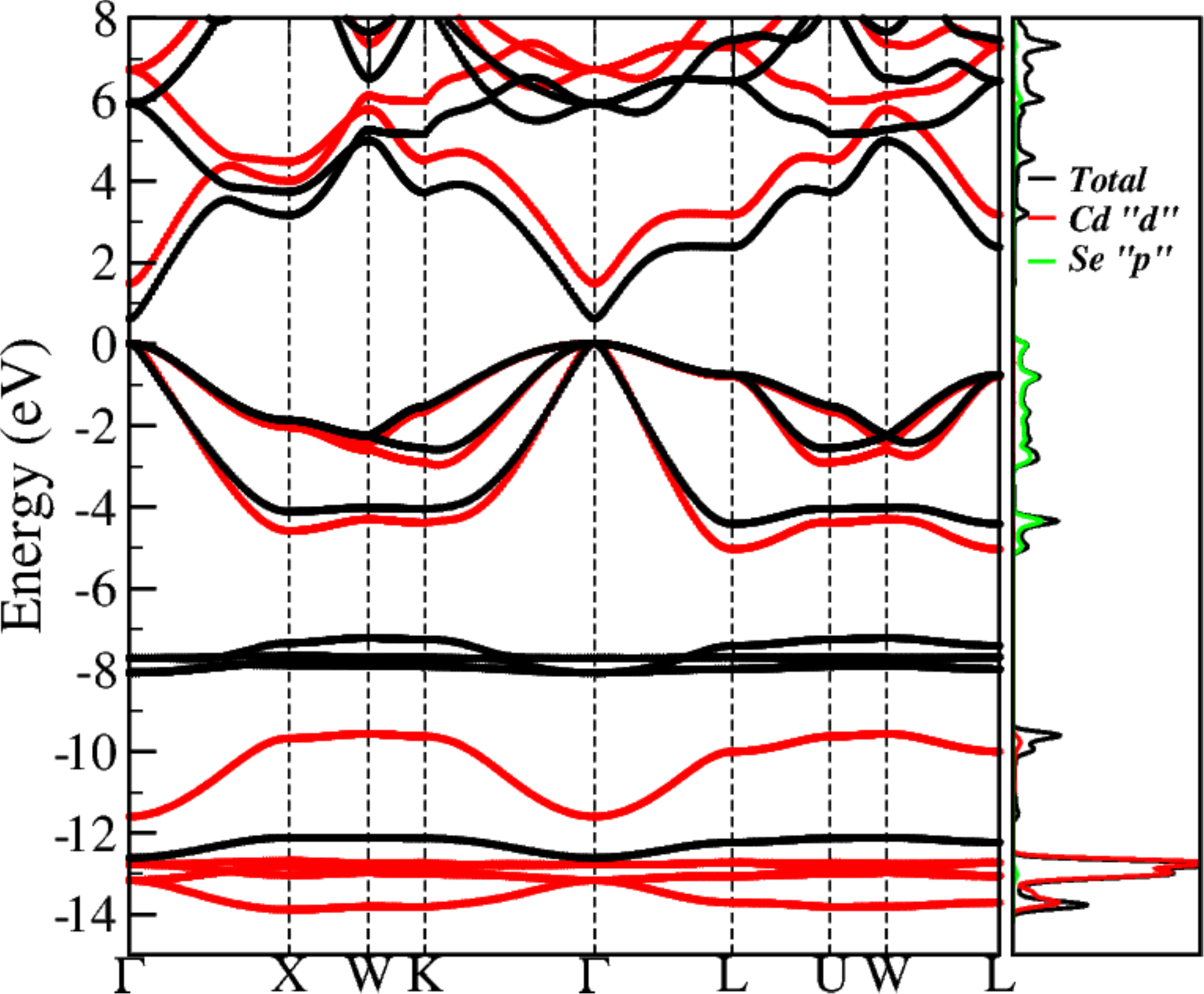}& \includegraphics[scale=0.25]{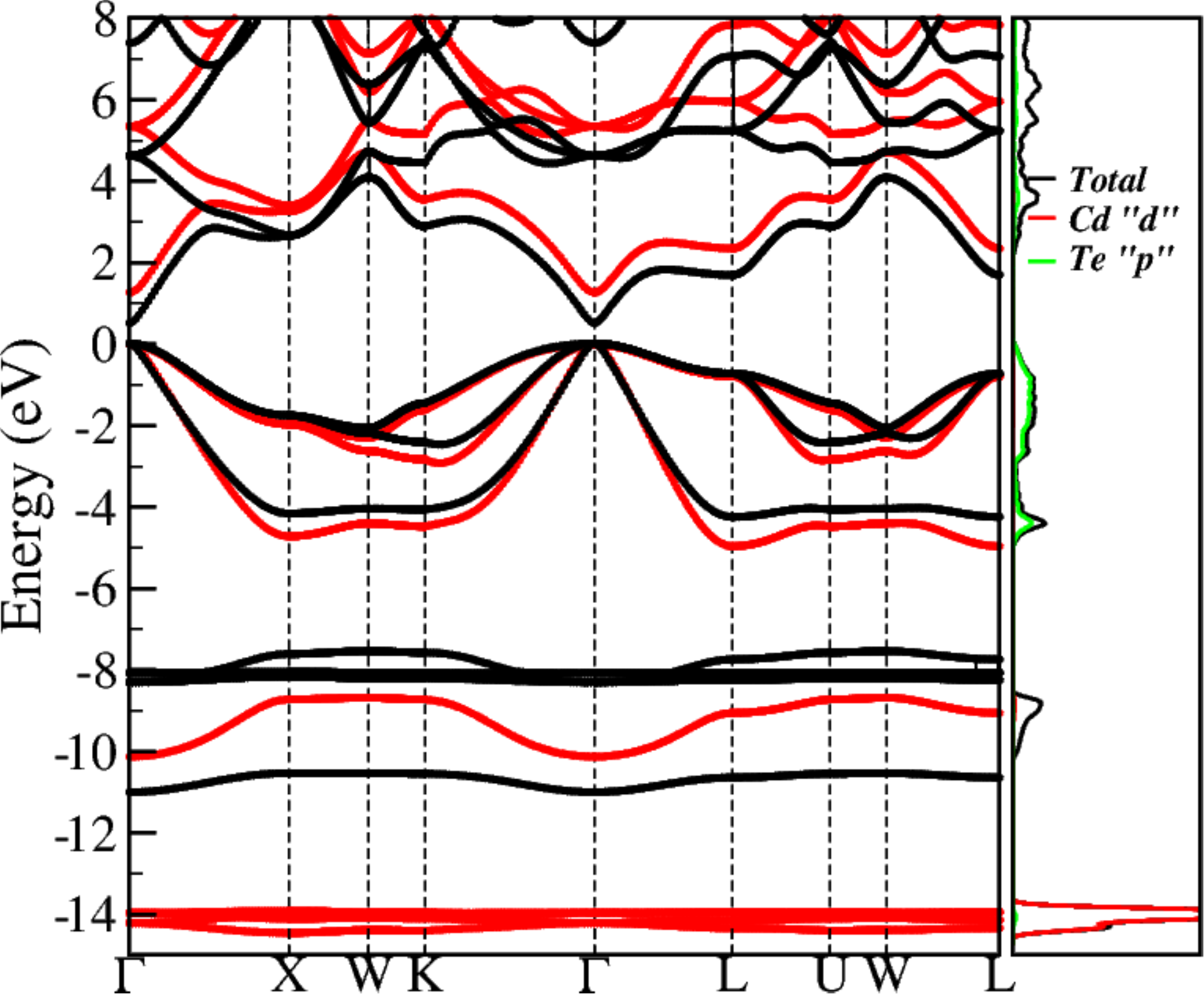} \\
CdO&CdS & CdSe &CdTe \\
\end{tabular}
\end{figure*}

\subsection{Lattice dynamical properties}
\label{phonon}

In the wide band gap semiconductors, the incorrect description of the covalency of the cation-anion bond affects the 
electrostatic properties of the system and, consequently, the phonon distribution and the coupling with the external fields.
This problem is more pronounced in the oxides compared to the chalcogenides. In ZnO the underestimated band-gap within PBE functionals leads to an overestimate of the high-frequency dielectric constant (${\varepsilon}^{\infty}$) and, in turn, a large discrepancy in the LO-TO splitting. The PBE value  of 5.24 is higher compared to the experimental value of 3.14 (Ref. \onlinecite{Nardelli_Scientific_Reports_2013}).
Similarly, in CdO the PBE dielectric constant is 7.13 compared to the experimental value of 5.3. This trend of overestimation of the high-frequency dielectric constant is observed in the other chalcogenides as well. 

The ACBN0 functional provides a
proper description of the electronic and structural features of the II-VI semiconductors and improves the dielectric and vibrational properties leading, in most cases, to smaller deviations with respect to the experimental data. 
The calculated high frequency
dielectric constants (${\varepsilon}_{\infty}$), the Born effective charges ($Z^{\star}$) and the zone-center phonon frequencies are summarized 
in Tables ~\ref{phonon-dielectric} and \ref{phonon-freq}. We did not find any references to vibrational properties calculated within the HSE method which is computationally very expensive especially for calculating the response functions.

\begin{figure*}
\caption{\label{fig:ZnX-phonon}(Color online)  Phonon dispersion of Zn\textit{X} (\textit{X}=S,Se,Te) within the ACBN0  method for the  \textit{zb} phase (red). The PBE phonon dispersions (black) is also shown for comparison. Open blue diamonds represent inelastic scattering data from Ref. \onlinecite{hennion-znseph-PhysLett71,zns-vagelatos1974-JCP}. We show only the directions in the Brillouin zone that have been measured, see the Supplemental Information for the full phonon dispersions.}
\begin{tabular}{ccc}
 \includegraphics[scale=0.3]{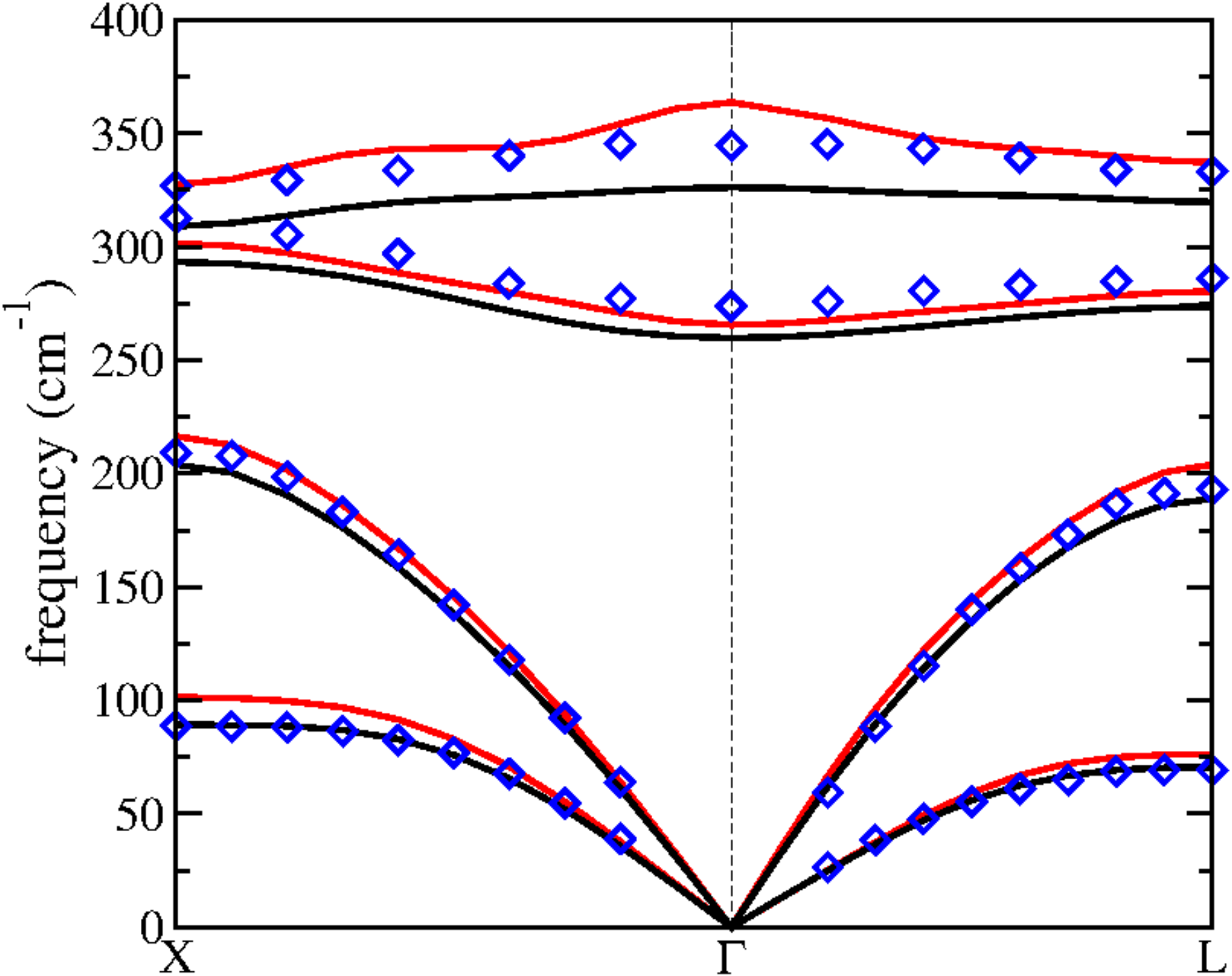} & \includegraphics[scale=0.30]{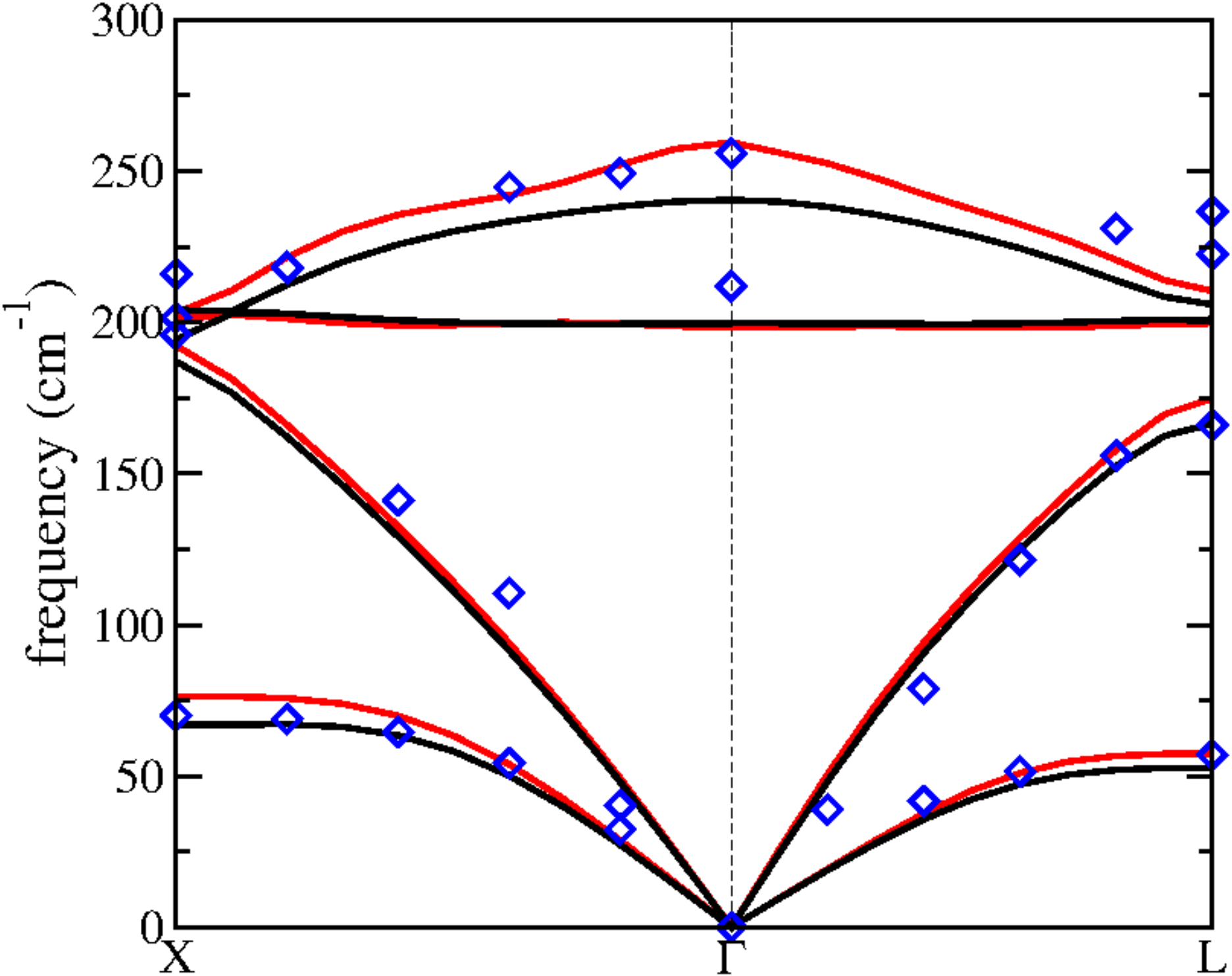}& \includegraphics[scale=0.30]{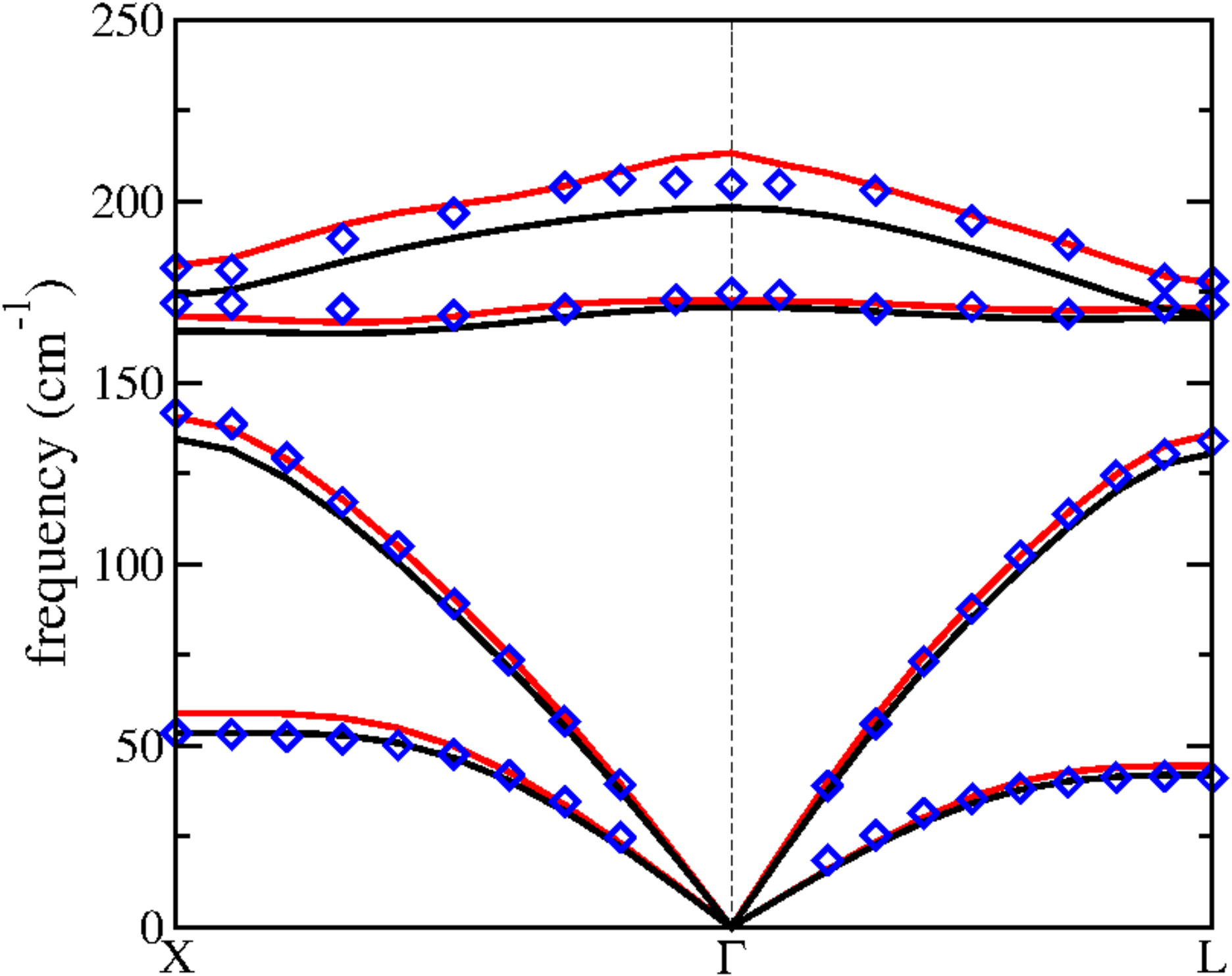} \\
 ZnS & ZnSe &ZnTe \\
\end{tabular}
\end{figure*}

\begin{figure*}
\caption{\label{fig:CdX-phonon}(Color online)  Calculated phonon dispersion of Cd\textit{X} (\textit{X}=O,S,Te) within the ACBN0 (red lines) and PBE (black lines) method for the \textit{zb} phase. Open blue diamonds represent experiment inelastic neutron scattering data are from Ref. \onlinecite{Oliva-CdOph-JAP13}~\onlinecite{cdte-rowe1974-ph-PRB}.We show only the directions in the Brillouin zone that have been measured, see the Supplemental Information for the full phonon dispersions.}
\begin{tabular}{ccc}
 \includegraphics[scale=0.30]{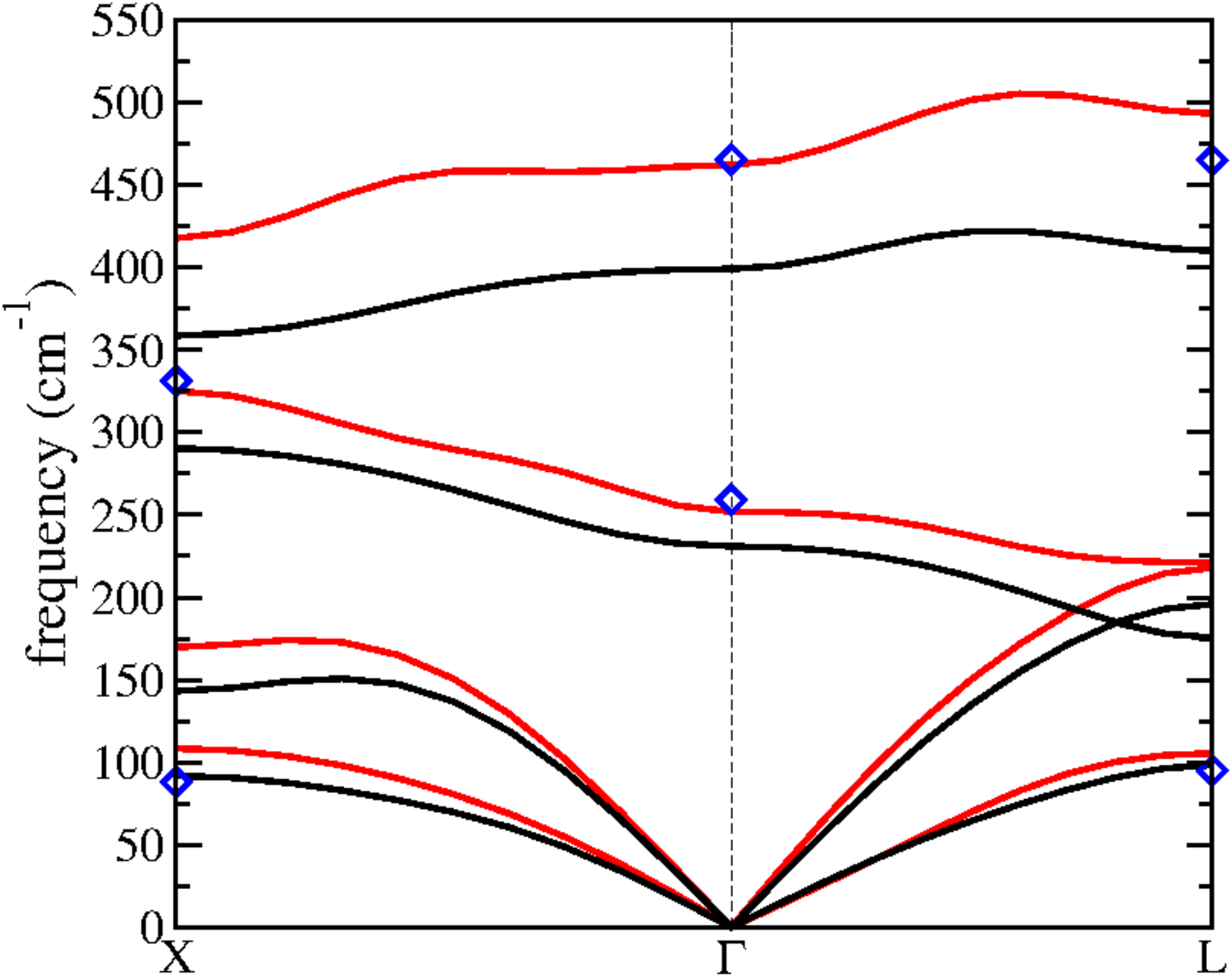} & \includegraphics[scale=0.30]{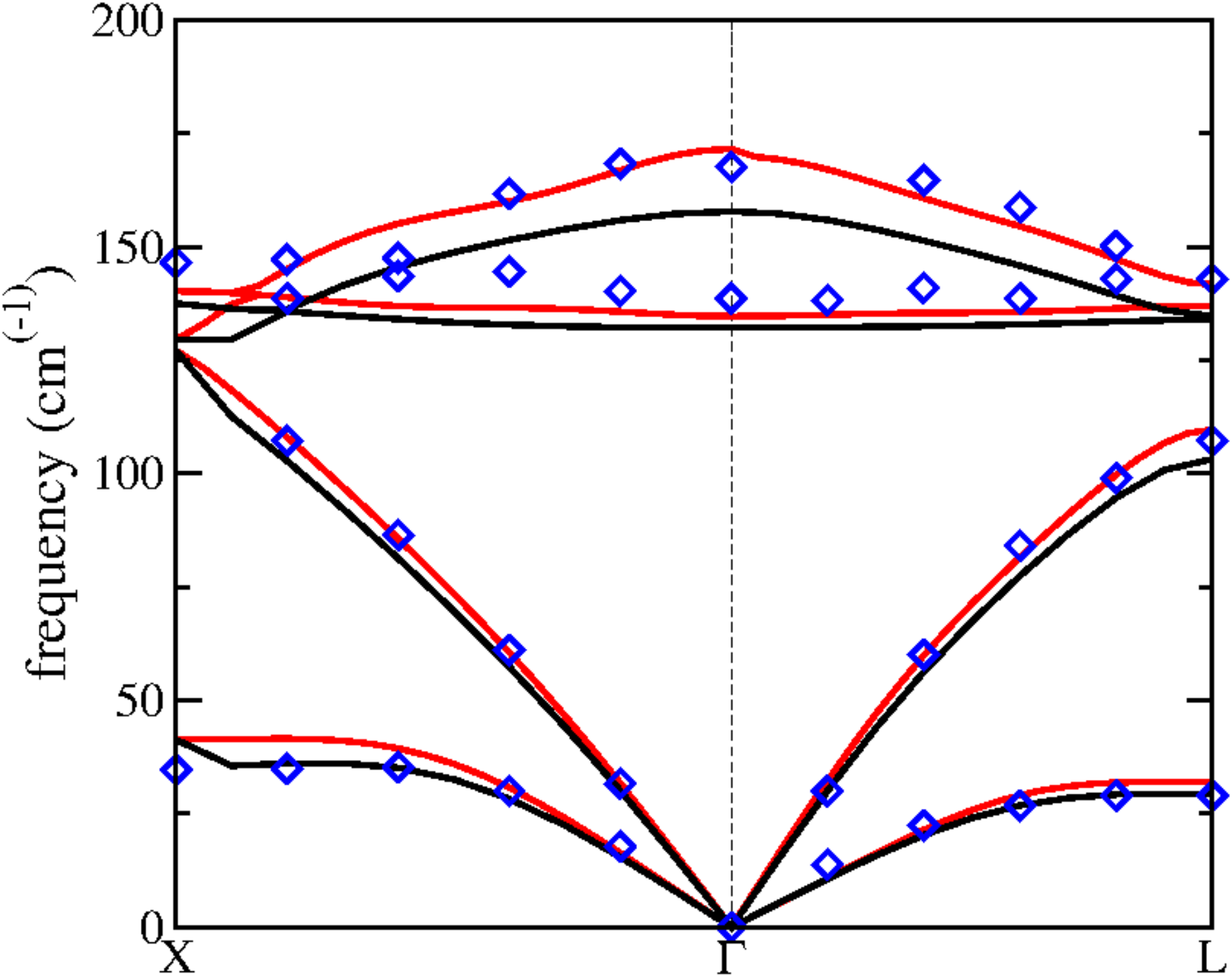}& \includegraphics[scale=0.30]{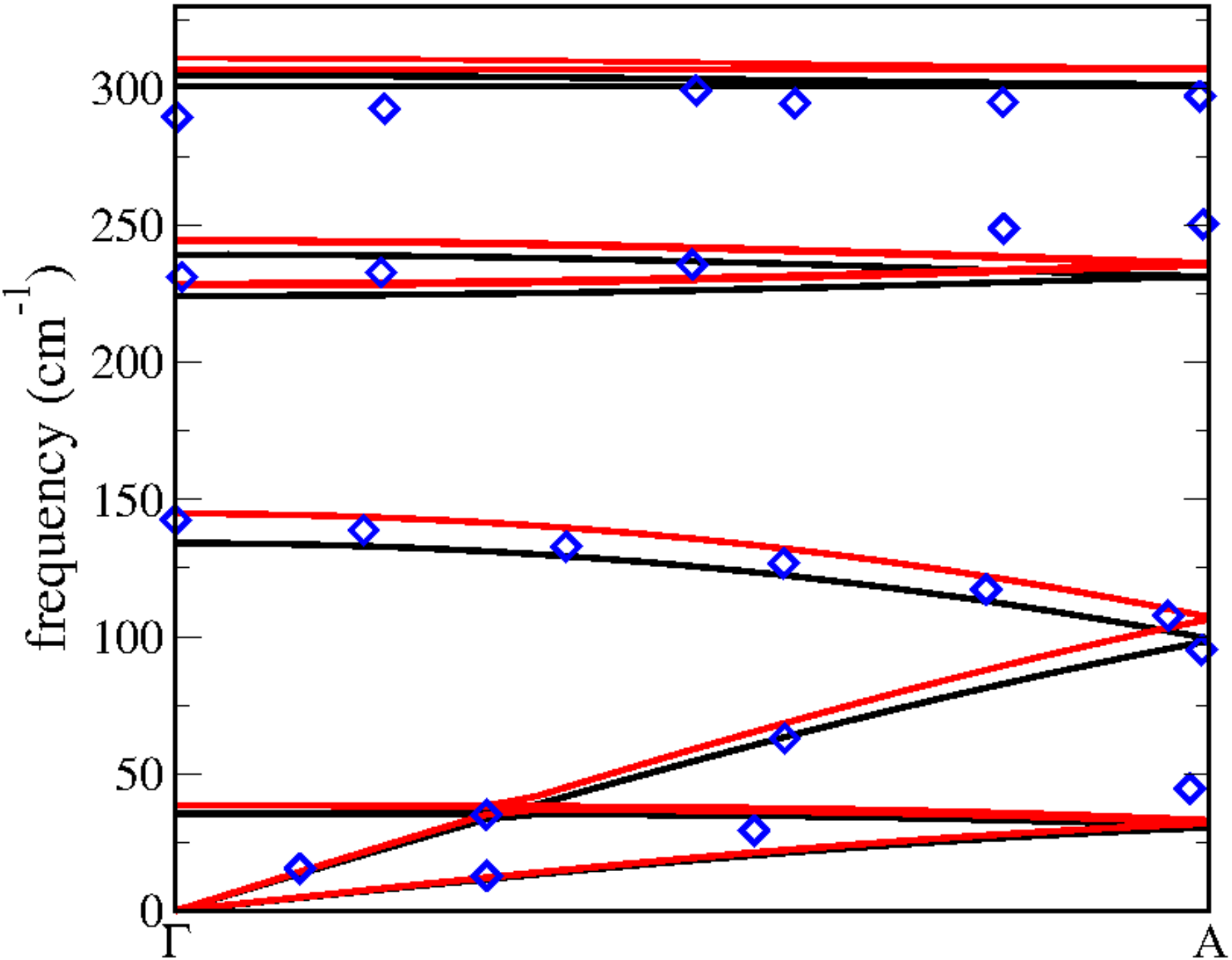} \\
 CdO& CdTe & CdS \\
\end{tabular}
\end{figure*}

In this section we also show the accuracy of the phonon spectrum for six representative compounds: ZnS, ZnSe, ZnTe, CdO, CdS, and CdTe. ACBN0 phonons frequencies are compared with the PBE and experimental values.
The dielectric properties and the vibrational spectrum of the II-VI semiconductors were calculated using a coupled finite-fields/finite differences approach as discussed extensively in Ref. \onlinecite{Nardelli_Scientific_Reports_2013}. The PBE and ACBN0 phonon dispersions are reported in Figures~\ref{fig:ZnX-phonon} and ~\ref{fig:CdX-phonon}. In the Zn chalcogenides the acoustic manifold is equally well reproduced within the PBE and the ACBN0 formalisms (deviations of few cm$^{-1}$ with respect to the available experimental data). For the optic manifold, however, ACBN0 improves the agreement with experiment. This result may be rationalized in terms of the more accurate description of the bonding.
In both CdO and CdTe, the ACBN0 values are in excellent agreement with the experiments and improve the PBE description especially for the optic manifolds. In CdS both ACBN0 and PBE are in good agreement with the experiments.


\begin{table}
\caption{\label{phonon-freq}Calculated frequencies of optic and acoustic phonons at specific high-symmetry points in the Brillouin zone. We report result computed with ACBN0 and PBE functionals for all the II-VI semiconductors.Comparison with experiments is also reported wherever available. Experimental data are from Refs.,\onlinecite{cdte-rowe1974-ph-PRB}(CdTe),\onlinecite{Oliva-CdOph-JAP13}(CdO),\onlinecite{hennion-znseph-PhysLett71}(ZnSe,ZnS)}
\begin{tabular}{llccc}
\hline
\hline
System & Mode &  PBE & ACBN0 & Experiment \\
\hline 
ZnS &TO($\Gamma$)  &259  & 266  &  273 \\
    &LO($\Gamma$)  &329  & 364 & 344 \\
    &TO(X)  &293  &301  & 312 \\
    &LO(X)  &309  &327  & 326 \\
    &TO(L)  &273  & 280 &  286 \\
    &LO(L) &319  & 337 & 333  \\
ZnSe &TO($\Gamma$)  &199     & 198 & 213 \\
    &LO($\Gamma$)  & 240   & 260 & 254  \\
    &TO(X)  & 204 &   203 & 216 \\
    &LO(X)  & 204  & 203& 225\\
    &TO(L)  &201  &   200& 222\\
    &LO(L) & 205  &   210& 236  \\ 
ZnTe &TO($\Gamma$)  &170  & 172 & 175 \\
    &LO($\Gamma$)  & 198 & 213 & 204 \\
    &TO(X)  &164 & 168 & 183 \\
    &LO(X)  &174  & 182 & 181  \\
    &TO(L)  & 167 & 170 & 171 \\
    &LO(L) & 168 & 177 & 177  \\   
CdO &TO($\Gamma$)  & 234  & 251  & 266 \\
    &LO($\Gamma$)  & 410  & 462 & 465  \\
    &TO(X) & 290 & 325 & 331 \\
CdTe &TO($\Gamma$) & 132 & 134 & 138 \\
    &LO($\Gamma$)  & 158 & 171 & 167  \\
    &TO(X)  &137  & 140 &\\
    &LO(X)  & 137 &140  &  147  \\
    &TO(L)  & 133 & 137 &  \\
    &LO(L) & 134 & 141 & 142  \\ 
 \hline   

\end{tabular}
\end{table}

\begin{table}
\caption{\label{phonon-dielectric}High frequency (${\varepsilon}_{\infty}$) dielectric constant and Born effective charges (Z) of the II-VI compounds computed with ACBN0 and PBE. The experimental data are from Ref. \onlinecite{Oliva-CdOph-JAP13,hennion-znseph-PhysLett71,zns-vagelatos1974-JCP,cdte-rowe1974-ph-PRB}.}
\begin{tabular}{lcccccc}
\hline
\hline
&\multicolumn{3}{c}{${\varepsilon}^{\infty}$} &\multicolumn{3}{c}{Z} \\
\hline
 System & PBE &ACBN0 & Experiment& PBE & ACBN0 & Experiment \\
\hline

ZnS &  5.98  &  4.12 & 5.13  & 2.02 & 2.10 & 2.1 \\
ZnSe &  7.40     &  4.98     & 5.70      & 2.12     & 2.17    & 2.21   \\
ZnTe &  9.38    & 6.01       &  7.2     & 2.22     & 2.19     &  2.0    \\
CdO  & 7.13      &  3.52     & 5.3      & 2.40     & 1.97     &     \\
CdTe &  9.13     & 5.56      & 7.1      & 2.45     & 2.34     & 2.06     \\
\hline
\end{tabular}
\end{table}

\section{Conclusions}
The procedure to compute from first principles the Hubbard correction as implemented in the ACBN0 functional largely improves the electronic structure with respect to PBE. The energy band gap in Cd and Zn oxides and chalcogenides computed within the ACBN0 formalism is very close to the value computed with hybrid functional at a computational cost of roughly 10\%. The ACBN0 functional generally improves the structural and the vibrational parameters reaching improved agreement with the experiments. Both hybrid functionals and ACBN0 have limited predictive value when considering the the position of the occupied \textit{d} manifolds, at least in ZnS, ZnSe, ZnTe, ZnO, CdS, CdSe, CdTe, and CdO that we investigated.

\begin{acknowledgments}
We thank A. Calzolari, H. Shi, I. Takeuchi, G. Hart and R. Forcade for various technical discussions that have contributed to the results reported in this article.

This work was supported by ONR-MURI under contract N00014-13-1-0635, DOD-ONR (N00014-14-1-0526) and the Duke University Center for Materials Genomics. 
S.C. acknowledges partial support by DOE (DE-AC02-05CH11231, BES program under Grant \#EDCBEE).
We also acknowledge the Texas Advanced Computing Center (TACC) at the University of Texas Austin for providing HPC resources, 
 and the CRAY corporation for computational assistance.
 \end{acknowledgments} 

\bibliographystyle{apsrev4-1}
\bibliography{xstefano4} 
\end{document}